%% file: main.tex
\documentclass[journal]{IEEEtran}

\input{config/packages}

\begin{document}

\input{config/title-page}

\input{common/abstract}

\input{sections/introduction}

\input{sections/related-works}

\input{sections/system-model}

\input{sections/proposed-system}

\input{sections/MetaGraphLoc}
\input{sections/performance-eval}
\input{sections/conclusion}


\balance

\bibliographystyle{IEEEtran}
\bibliography{references,related-works}

\end{document}

%% file: config/packages.tex
\usepackage{amsmath,amsfonts}
\usepackage{array}
\usepackage{subcaption}
\usepackage{textcomp}
\usepackage{stfloats}
\usepackage{url}
\usepackage{verbatim}
\usepackage{graphicx}
\usepackage{cite}
\usepackage{amsthm}

\usepackage{soul}
\usepackage{cuted}
\usepackage[linesnumbered,ruled,vlined]{algorithm2e}
\usepackage{balance}
\usepackage{tabularx}
\usepackage{makecell}
\usepackage{multirow}
\usepackage{booktabs} 

\usepackage{amsmath,amsfonts}
\usepackage{algorithmic}
\usepackage{array}
\usepackage{graphicx}
\usepackage{cite}
\usepackage{url}
\usepackage{xcolor}
\usepackage{hyperref} 
\usepackage{blindtext}

%% file: config/title-page.tex
\title{MetaGraphLoc: A Graph-based Meta-learning Scheme for Indoor Localization via Sensor Fusion}


\author{
	\IEEEauthorblockN{Yaya Etiabi, Eslam Eldeeb, Mohammad Shehab, Wafa Njima, Hirley Alves, Mohamed-Slim Alouini,  \mbox{and El Mehdi Amhoud} \\
	}
	\thanks{Yaya Etiabi and El Mehdi Amhoud are with College of Computing, Mohammed VI Polytechnic University, Benguerir, Morocco (email: yaya.etiabi@um6p.ma, elmehdi.amhoud@um6p.ma).}
 \thanks{Wafa Njima is with LISITE Laboratory, Institut Supérieur d'Électronique de Paris, Paris, France (e-mail: wafa.njima@isep.fr).}
 \thanks{Eslam Eldeeb and Hirley Alves are with Centre for Wireless Communications (CWC), University of Oulu, Finland. (e-mail: eslam.eldeeb@oulu.fi, hirley.alves@oulu.fi).} 
\thanks{Mohammad Shehab and Mohamed-Slim Alouini are with CEMSE Division, King Abdullah University of Science and Technology (KAUST), Thuwal 23955-6900, Saudi Arabia (email: mohammad.shehab@kaust.edu.sa, slim.alouini@kaust.edu.sa).}
\thanks{The work of E. Eldeeb and H. Alves was partially supported by the Research Council of Finland (former Academy of Finland) 6G Flagship Programme (Grant Number: 346208) and by the European Commission through the Hexa-X-II (GA no. 101095759). 
        }
}

\maketitle

%% file: common/abstract.tex
\begin{abstract}
Accurate indoor localization remains challenging due to variations in wireless signal environments and limited data availability. This paper introduces MetaGraphLoc, a novel system leveraging sensor fusion, graph neural networks (GNNs), and meta-learning to overcome these limitations.
MetaGraphLoc integrates received signal strength indicator measurements with inertial measurement unit data to enhance localization accuracy. Our proposed GNN architecture, featuring dynamic edge construction (DEC), captures the spatial relationships between access points and underlying data patterns. MetaGraphLoc employs a meta-learning framework to adapt the GNN model to new environments with minimal data collection, significantly reducing calibration efforts.
Extensive evaluations demonstrate the effectiveness of MetaGraphLoc. Data fusion reduces localization error by 15.92\%, underscoring its importance. The GNN with DEC outperforms traditional deep neural networks by up to 30.89\%, considering accuracy. Furthermore, the meta-learning approach enables efficient adaptation to new environments, minimizing data collection requirements.
These advancements position MetaGraphLoc as a promising solution for indoor localization, paving the way for improved navigation and location-based services in the ever-evolving Internet of Things networks.
\end{abstract}

\begin{IEEEkeywords}
Indoor localization, meta-learning, graph neural networks, data fusion.
\end{IEEEkeywords}

%% file: sections/introduction.tex
\section{Introduction}
\IEEEPARstart{R}{ecent} advances in machine learning and wireless communication technologies have significantly accelerated the development of indoor localization systems. These systems are increasingly integral to various critical Internet of Things (IoT) applications. Indoor localization, unlike its outdoor counterpart, faces substantial challenges due to the complex nature of indoor environments, where satellite-based positioning systems like GPS are unreliable \cite{IPS_survey}. These challenges are characterized by signal attenuation, multi-path propagation, and the need to navigate through various obstructions. As a result, there has been a growing reliance on technologies such as Wi-Fi, Bluetooth Low Energy (BLE), Radio Frequency Identification (RFID), and inertial measurement units (IMUs) \cite{next}, each of which offers unique methods for estimating indoor positions. 
However, these methods alone often fall short of the precision required for advanced applications. Indeed,
the applications of indoor localization are vast, encompassing asset tracking, navigation in large facilities like airports and shopping malls, and enhancing safety protocols during emergencies. 
As these technologies evolve, the demand for increased accuracy and reliability in indoor localization systems continues to grow, driving innovation and expanding the potential for seamless indoor navigation and advanced location-based services.

In response to this growing demand, machine learning (ML) has emerged as a powerful tool for processing the heterogeneous and high-dimensional data generated by these localization technologies. This data-driven approach enables the extraction of accurate positioning information from complex indoor environments.
In this context, many research papers have been targeting the indoor localization problem via ML \cite{ML}. For instance, the work in \cite{Etiabi2023FederatedLB} proposed a radio signal strength indicator (RSSI)-aided federated learning approach for indoor localization using deep neural networks. The authors of \cite{Cross} suggested a cross-environment deep learning approach to address the problem of insufficient labeled measurements using a joint semi-supervised and transfer learning algorithm. In \cite{Elesawi}, Elesawi et al. applied a recurrent neural network (RNN) scheme that depends on Wi-Fi fingerprinting to tame the issue of data pre/post-processing and parameter tuning. Meanwhile, \cite{SF} considers the spreading factor and applies Deep reinforcement learning to improve localization accuracy. More recently, the work in \cite{TURGUT2024196} applies long-short-term temory (LSTM) to capture long-term dependencies between the signal features and a convolutional neural network (CNN) to extract local spatial signal patterns. The results of this combination were promising in terms of localization accuracy. 

While neural network (NN)-based approaches offer high accuracy, their training often requires large datasets and significant computational resources. Federated learning (FL) has emerged as a potential solution to address these limitations as well as privacy concerns. In  \cite{FD}, the authors explore federated distillation (FD), a prominent FL technique. Their findings demonstrate that FD can achieve reasonable accuracy with lower communication overhead and energy consumption than traditional FL. Additionally, graph neural networks (GNNs) have emerged as a specialized class of neural networks designed to address problems related to graph interpretations, such as geometric correlation scenarios, further expanding the toolkit available for enhancing indoor localization systems. For example, the work in \cite{GNN_Chinese} applied GNNs in large-scale 2D localization as a proof of concept. The authors of \cite{percision} developed a GNN-based localization scheme using signal RSSI and CSI fingerprints. Their proposed scheme was characterized by high precision and a localization error down to 0.17 m. In \cite{few_shot},  transfer learning was applied with GNN. The combined scheme achieved indoor localization using a small amount of labeled data. Moreover, the work in \cite{Yu} utilized the CSI with a federated complex-valued NN, which achieved better positioning accuracy compared to real-valued NN. Meanwhile, the work in \cite{CSI} exploited amplitude and phase information in GNN-based localization.

Despite these advancements, there remain challenges in effectively fusing data to enhance location information accuracy for robust models and adapting these models to unseen environments, which often requires extensive data collection and impairs user privacy. To address these challenges, we introduce MetaGraphLoc, a novel framework that leverages GNN-based data fusion to create a comprehensive feature representation that integrates RSSI and IMU data. Furthermore, MetaGraphLoc incorporates a meta-learning component that enables the model to learn from diverse environments, facilitating rapid adaptation to new, unforeseen indoor environements.

The overall contributions of this paper can be summarized as follows:
\begin{itemize}
    \item \textbf{Data fusion for enhanced localization accuracy:} we integrate WiFi RSSI measurements with IMU data to create a comprehensive input for indoor localization. This fusion significantly reduces localization error by 15.92\%, highlighting the importance of combining multiple data sources to improve accuracy and robustness in varying indoor environments.
    \item \textbf{GNN architecture with dynamic edge construction:} We propose a novel GNN architecture that utilizes DEC to effectively capture spatial relationships between access points and underlying data patterns. This approach allows the GNN to adjust its structure based on the data dynamically, leading to superior performance. Our GNN model outperforms traditional deep neural networks (DNNs) by up to 30.89\% in localization accuracy, demonstrating the effectiveness of graph-based methods in handling the challenges of indoor localization.
    \item \textbf{Meta-learning framework for fast adaptation:} We develop a meta-learning framework that enables the GNN model to quickly adapt to new environments with minimal data collection and calibration. This framework addresses the dynamic behaviour of indoor environments, ensuring that the model remains effective even when deployed in previously unseen settings. The meta-learning approach reduces the hunger for extensive data collection, making it practical for real-world applications where rapid deployment and adaptability are crucial.
\end{itemize}

The rest of the paper is organized as follows: Section~\ref{sec:relworks} discussed the related works and Section~\ref{sec:sysm} describes the system model and defines the problem. Next, Section\ref{sec:graphloc} presents the data fusion and the GNN model, while Section \ref{sec:metagraphloc} depicts the proposed MetaGraphLoc architecture. Section~\ref{sec:perfeval} elucidates the experimental results, while Section~\ref{sec:conclusion} concludes the paper.

%% file: sections/related-works.tex
\section{Related works}
\label{sec:relworks}

\subsection{Data Fusion for Indoor Localization}
%
Traditional indoor localization methods often rely on single sensor modalities but struggle with environmental variations and signal instability. To address these challenges, recent research has explored data fusion techniques, combining multiple sensors to enhance localization accuracy.

For instance, \cite{Zou2017AccurateIL} leveraged IMU readings, WiFi RSSI measurements, along with opportunistic iBeacon corrections using a particle filter. It employs a pedestrian dead reckoning (PDR) approach to estimate the user's walking distance and direction based on IMU data collected from a smartphone. The estimated position from PDR is then fused with the position obtained through WiFi fingerprinting to mitigate drift errors inherent in fingerprinting. Similarly, \cite{Herath2021FusionDHLWI} proposed a multi-modal sensor fusion method that combines WiFi, IMU, and floorplan information to infer location history in indoor environments. This approach leverages the spatial constraints provided by the floorplan to achieve more precise localization compared to sensor data alone. However, a major limitation of this work is the requirement for floorplan information, which may not be readily available in all indoor environments, hindering its widespread applicability. Therefore, more recent research focuses on effectively fusing data from readily available sensors, such as WiFi and IMU, to bypass the limitations of requiring floorplan information. For example, \cite{e2eseq} constructs a dataset containing WiFi RSSI and IMU readings specifically for training machine learning models. They then propose an end-to-end neural network-based localization system that utilizes both RSSI measurements and IMU data to estimate user locations. A more comprehensive exploration is provided in \cite{survey} for systems that adopt measurements of, for instance, signal time-of-arrival (ToA), time-difference-of-arrival (TDoA), and angle-of-arrival (AoA). 
%

%
These studies show that data fusion is a promising approach for indoor localization. However, blindly using deep learning for fusion can yield unpredictable results due to its "black box" nature, as seen in \cite{9746071}, where fusion didn’t significantly improve performance. To address this, this work introduces a graph-learning-based framework that efficiently fuses RSSI and IMU data, leveraging the wireless network's inherent structure to enhance localization accuracy.

\subsection{Graph Neural Networks for Indoor Localization}

%
GNNs have become a powerful tool for indoor localization, particularly due to their ability to model spatial relationships between data points like access points (APs) or reference points (RPs). In \cite{Sun2021ANG}, the authors demonstrated the effectiveness of using graph convolution networks (GCNs) in fingerprint-based localization systems by leveraging the geometric relationships between APs. Their proposed scheme achieves high accuracy by fully utilizing the GCN architecture's feature extraction capabilities.

Moreover, in \cite{Zhao2021GraphIPSCA}, the authors took a different approach by using reference points for graph construction.
Their system leverages crowd-sourced WiFi and IMU data from smartphones to dynamically generate radio maps, eliminating the need for calibration or pre-existing maps. The algorithm fuses the crowd-sourced data into a graph-based formulation and then employs multi-dimensional scaling to compute user location based on their step patterns. 

Building on the same concept, \cite{factorgraphs} introduced the use of factor graphs with multimodal localization data, including ultrawideband (UWB) ranging data, WiFi RSSI, IMU, and floorplan information. They proposed a framework that integrates ranging and fingerprinting using factor graphs, achieving a favorable balance between accuracy and deployment cost. 

Numerous other works, including \cite{10085729, 10105890, FactorGraph-Trans}, have explored the use of GNNs for indoor localization by employing either reference points (RPs) or APs for graph formulation, and investigating both individual sensor modalities and data fusion techniques.
As a result, the benefits of graph-based indoor localization are well-established, though this approach incurs significant computational costs. These costs limit the development of GNN-based indoor localization, as training GNNs is resource-intensive, and data collection is costly.
In our work, after developing a baseline GNN-based data fusion approach for indoor localization, we propose a meta-learning framework that allows for quick adaptation to new indoor environments with reduced data collection and computational complexity.

\subsection{Meta-Learning for Indoor Localization}

Meta-learning algorithms learn from a collection of tasks, allowing them to quickly adapt to unseen environments that share similar characteristics with minimal data \cite{maml}. As demonstrated in \cite{femloc}, this approach is particularly beneficial for indoor localization, where extensive data collection for each unique environment is often impractical. In this context, meta-learning treats each unique localization problem with its corresponding data as a separate task. A meta-model is then trained on a collection of such tasks, allowing it to initialize new models for unseen environments with just few data samples, facilitating rapid adaptation.

For example, \cite{Wei2023AMA} demonstrates the generalizability of meta-learning for indoor localization. Their work utlizies previously collected channel state information (CSI) fingerprints from various environments. This data is used to train a meta-learning model that acts as an effective "initializer" for new localization tasks. This initializer allows a new model to achieve high performance in a new environment with only a small amount of additional labeled training data. Similarly, in \cite{Gao2022MetaLocLT-conf} and its extended version \cite{Gao2022MetaLocLT-journal}, the authors proposed a solution applicable to various data-driven wireless localization problems. Their work eploits wireless signal data (both RSSI and CSI) previously collected from various well-calibrated historical environments. This data is used in a two-stage training process to obtain meta-parameters that serve as the initialization for neural networks deployed in new environments. This approach enables fast adaptation to unseen locations and can be applied to problems formulated as either regression or classification tasks. Other works, such as \cite{Owfi2023AMB, Wei2023AMA}, also investigated the prominence of meta-learning for indoor localization tasks, highlighting the benefits of this technique and making it increasingly appealing to the indoor localization research community.

While existing research explores separately data fusion, GNNs, and meta-learning for indoor localization, our proposed system, MetaGraphLoc offers a novel approach that combines these techniques to achieve superior performance. Our framework leverages data fusion to create a comprehensive feature representation that integrates RSSI and IMU data. This enriched data is utilized within a GNN architecture specifically designed for indoor localization tasks. Furthermore, MetaGraphLoc incorporates a meta-learning component that enables the model to learn from a set of diverse environments, facilitating fast adaptation to new unseen locations. This combination of data fusion, GNNs, and meta-learning within MetaGraphLoc sets it apart from existing approaches and contributes to significant improvements in accuracy, efficiency, and adaptability for indoor localization tasks.

%% file: sections/system-model.tex
\section{System model and problem formulation}
\label{sec:sysm}
\begin{figure}[!t]
    \centering
    \includegraphics[width=\linewidth]{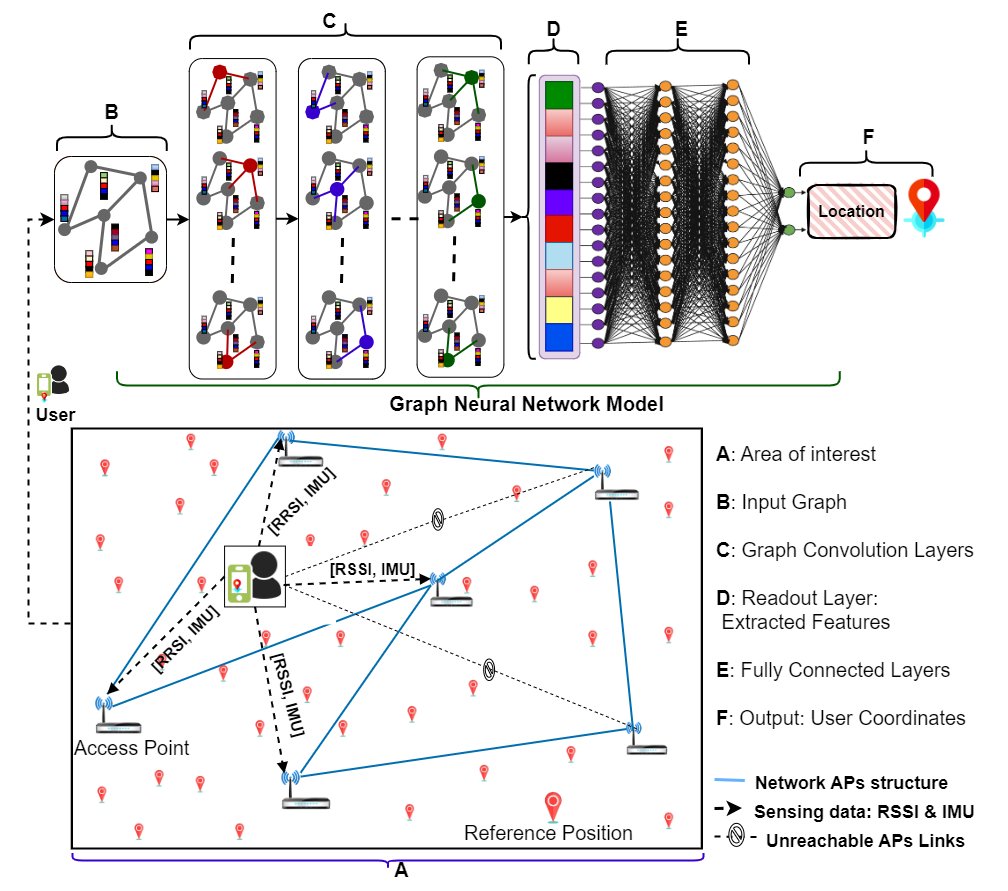}
    \caption{System model of the AP-based graph formulation for indoor localization. }
    \label{fig:sysm}
    \vspace{-0.2in}
\end{figure}
This study explores wireless localization in an indoor IoT network scenario. When an IoT device enters this environment, potentially introduced by a user, it can sense the wireless access points (WAPs), which provide information on its position. Additionally, the IoT device may be equipped with additional sensors, such as the IMU, that can be used to complement the location information. In this work, we consider an indoor localization system that leverages both WIFI RSSI and IMU readings from the onboard sensors of the IoT devices to provide location awareness. Furthermore, the WIFI RSSI measurements are highly influenced by the environment layout, the deployment strategy, and some additional mutual information of the APs. To improve localization accuracy, we aim to incorporate this extra information through graph learning techniques. Specifically, we introduce graph learning techniques for extracting features based on the network’s geometry, enriched with data from the IMU sensors, as depicted in Fig. ~\ref{fig:sysm}.

We consider $M$ WiFi access points deployed within the indoor environment. Let \mbox{$\mathcal{W}_{AP} = \left[ AP_0, AP_1, ..., AP_M \right]$} be the set of APs within the environment. When an IoT device enters the environment, it scans it for the available APs. This results in the RSSI measurement vector \mbox{$\boldsymbol{\gamma}_\ell = \left[ RSSI_{\ell,0}, RSSI_{\ell,1}, ..., RSSI_{\ell, M} \right]$}, where $\ell$ is the device's current location from which the measurements are done. Note that some APs may be out of range, and the resulting RSSI measurements are set to arbitrary values (e.g., placeholder) to indicate the missing measurements. When the positions of the APs are known and denoted as \mbox{$\boldsymbol{x}_{AP} = \{x_m\}_{m=1,2,..., M}$}, the APs can furnish the location details of the IoT device through wireless propagation models. For instance, according to the log-distance path loss (LDPL) model, the received signal strength $RSSI_{d,m}$ is expressed as \mbox{$RSSI_{\ell,m} = P_{Rx} = P_{Tx} - PL_{\ell,m}$}, where $P_{Tx}$, $P_{Rx}$, and $PL_{\ell,m}$ are the transmitted power, the received power, and the path loss, respectively.
The path loss $PL_{\ell,m}$ is further defined as follows:
\begin{equation}
     PL_{\ell,m} = PL_0 + 10\beta \log _{10}\left(\frac {\left \| x_\ell - x_m \right \|}{D_{0}}\right)+X_{\ell,m}^\sigma,
     \label{equ:ldpl}
\end{equation}
with $\beta$ being the path loss exponent, $D_0$ the reference distance, and $X_\sigma$ representing the log-normal shadowing with standard deviation $\sigma$ in dB. $PL_0$ is the path loss at the reference distance $D_0$. From \eqref{equ:ldpl}, we see how the location information of the IoT device is incorporated in the RSSI measurements and can be determined using some probabilistic model, namely the likelihood function defined as in \eqref{equ:lkhood} at the bottom of the page.
\begin{strip}
    \hrulefill
    \begin{equation}
p\left(RSSI_{\ell, m} \mid x_\ell, x_m\right)= \frac{1}{\sqrt{2 \pi} \sigma} \exp \bigg\{\frac{-1}{2 \sigma^2}\bigg[P_{Tx}-PL_0
 -10 \beta\lg \left( \frac{\left\|x_\ell-x_m\right\|}{D_0}\right)-RSSI_{\ell, m}\bigg]^2\bigg\}.
\label{equ:lkhood} 
\end{equation}
\vspace{-0.1in}
\end{strip}

Besides the location information of the APs, this probabilistic approach requires knowledge of the variables of the propagation channel. This results in effort-intensive and model-based localization development which has shown some limitations in large-scale and dynamic environments. Rather than relying on such a probabilistic model, we opt for a data-driven approach by building prior knowledge of the environment through a series of measurements to serve as an informative dataset for the localization task.

To do so, we consider $N$ reference positions (RPs) within the environment denoted by \mbox{$\mathcal{P} =\left\{\mathcal{P}_n \right\}_{n=1,2,...,N}$}, and for each RP $\mathcal{P}_n$, the IoT device collects RSSI measurements from the $M$ APs and constructs the RSSI vector \mbox{$\boldsymbol{\gamma}_n = \left[ RSSI_{n,0}, RSSI_{n,1},\cdots, RSSI_{n,M} \right]$} such that \mbox{$\mathcal{P}_n \mapsto \boldsymbol{\gamma}_n$}. This results in $N$ radio fingerprint observations defined by \mbox{$\left\{ \mathcal{P}, \boldsymbol{\gamma} \right\}  = \left\{ \mathcal{P}_n \mapsto \boldsymbol{\gamma}_n \right\}_{n=1,2,..., N}$}, also known as the radio map of the environment. 


Moreover, at each RP, the IMU measurements from the device's onboard sensors are recorded in addition to the RSSI measurements obtained from the different APs.
Indeed, the IMU acts as a versatile sensor suite, typically comprising three key components: accelerometers, gyroscopes, and often, magnetometers. 
These IMU sensors complement WiFi RSSI measurements by providing a comprehensive picture of the device's motion and orientation within its environment. While IMUs excel at tracking movement (e.g., successive positions and predicting the next position), they lack absolute location information alone. The IMU complements the RSSI measurements within our graph formulation in this work, as detailed in Section \ref{sec:graphloc}.




%% file: sections/proposed-system.tex
\section{Data fusion and graph formulation}
\label{sec:graphloc}

\subsection{Data fusion}
 Let \mbox{$\boldsymbol{\gamma}_n = \boldsymbol X^{\text{RSSI}}_n  = \left[ RSSI_{n,0}, RSSI_{n,1},\cdots, RSSI_{n,M} \right]$}, where \(M\) is the number of APs. In addition, for the reference position \(\mathcal{P}_n\), we have an associated IMU feature vector \(\boldsymbol X^{\text{IMU}}\), representing IMU information at that location. 
 %
 
This information is fused with the RSSI measurements to provide more context to the localization system.
Consequently, the fused feature vector for each reference position is denoted as \( \boldsymbol X^{\text{fused}}\) defined as:
\begin{equation}
\label{eq:xfusion}
    \boldsymbol X^{\text{fused}} = \varphi(\boldsymbol X^{\text{RSSI}}, \boldsymbol X^{\text{IMU}}).
\end{equation}
Here, \(\varphi (\cdot)\) is a fusion function that combines information from both modalities. More specifically, \(\varphi\) aggregates the RSSI and IMU features and creates new features for more relevance to the localization. In this work, we employ graph learning to reproduce the behavior of this function, as described in the next sections.












\subsection{Graph formulation}
\label{sec:graphformulation}
Formally, a graph is denoted by \( \mathcal{G} = (\mathcal V, \mathcal E) \), where \( \mathcal V \) is the set of vertices, and \(\mathcal E \) is the set of edges. The vertices are also known as nodes or points, which represent the individual entities within the graph. In contrast, the edges represent the connections between vertices. They can be directed (indicating a one-way relationship) or undirected (indicating a two-way relationship). Additionally, edges can be weighted to represent the strength or cost of the connection between connected vertices.  Graphs are widely used in various computer science applications due to their versatility and ability to model complex relationships. In particular, we formulate the indoor localization problem using graphs in this work. Indeed, as depicted in Fig. \ref{fig:sysm}, given RSSI measurements from an IoT device at position $\mathcal{P}_n$ in the network, the resulting fingerprint vector can be regarded as a graph where each AP constitutes a node, and the connection between APs is determined by their spatial relationship. More specifically, we consider the graph \(\mathcal{G}_{\text{AP}} = \left( \mathcal{V}_{\text{AP}}, \mathcal{E}_{\text{AP}}\right)\), where \(\mathcal{V}_{\text{AP}} = \{\text{AP}_m \}_{m=1,2,...,M}\) is the set of APs  and $\mathcal{E}_{\text{AP}}$ the edges representing the connections between APs. 

The adjacency matrix describes if vertices are close to each other and can be defined as 
\(\mathcal{A}_{\text{\text{AP}}} = [a_{ij}]_{\forall_{ij}}\). We may construct an informative adjacency matrix to define the edges using the approaches listed next.  
\subsubsection{Using Statistical information}
This statistical information is derived from the fingerprint database. This database embodies prior knowledge of the radio environment, facilitating supervised learning of our localization model.
Each element \(a_{ij}\) of the adjacency matrix represents the connection strength between nodes \(\text{AP}_i\) and \(\text{AP}_j\) in the graph \(\mathcal{G}_{\text{AP}}\). This connection strength is derived from the statistical correlation or similarity between the RSSI fingerprints associated with \(\text{AP}_i\) and \(\text{AP}_j\). In other words, the adjacency matrix is a $M\times M$ matrix that encodes the spatial relationships between different APs based on the collected RSSI data and is defined by
\begin{equation}
    \mathbf{A}_{\text{AP}} = \begin{pmatrix}
a_{11} & a_{12} & \dots & a_{1M} \\
a_{21} & a_{22} & \dots & a_{2M} \\
\vdots & \vdots & \ddots & \vdots \\
a_{M1} & a_{M2} & \dots & a_{MM}
\end{pmatrix}.
\end{equation}
Using Pearson correlation coefficient as similarity measure, the elements $a_{ij}$ of $\mathbf{A}_{\text{AP}}$ can be calculated as:
\begin{equation}
     a_{ij} = \frac{\sum_{k=1}^{N} (\gamma_{ik} - \bar{\gamma}_i)(\gamma_{jk} - \bar{\gamma}_j)}{\sqrt{\sum_{k=1}^{N} (\gamma_{ik} - \bar{\gamma}_i)^2 \sum_{k=1}^{N} (\gamma_{jk} - \bar{\gamma}_j)^2}}.
\end{equation}
Here,
\( \gamma_{ik} \) and \( \gamma_{jk} \) are the RSSI values from Fingerprint \( k \) associated with \( \text{AP}_i \) and \( \text{AP}_j \) respectively.
\( \bar{\gamma}_i \) and \( \bar{\gamma}_j \) are the mean RSSI values for \( \text{AP}_i \) and \( \text{AP}_j \) respectively, and \( N \) is the total number of fingerprints. 

\subsubsection{Joint appearance likelihood}
Alternatively, we propose a second statistical information-based approach using the likelihood of detecting \( \text{AP}_i \) and \( \text{AP}_j \) at the same time. This represents how frequently \( \text{AP}_i \) and \( \text{AP}_j \) appear simultaneously through the entire fingerprint database.
It can be formulated  by defining $a_{ij}$ as follows:
\begin{equation}
    a_{ij} = \frac{\sum_{\boldsymbol \ell \in \{\mathcal P_n\}} P_i(\boldsymbol \ell) \cdot P_{j|i}(\boldsymbol \ell)} {\sum_{\boldsymbol \ell \in \{\mathcal P_n\}} P_i(\boldsymbol \ell)}.
\end{equation}
$P_{j|i}(\boldsymbol \ell)$ represents the conditional probability of encountering \( \text{AP}_j \) at location $\boldsymbol \ell$ given that \( \text{AP}_i \) is also detected at $\boldsymbol \ell$.
The summation iterates over all locations $\ell \in \{\mathcal P_n\}$, calculating the product of $P_i(\boldsymbol \ell)$ and $P_{j|i}(\boldsymbol \ell)$, and then dividing by the sum of $P_i(\boldsymbol \ell)$ across all locations. This provides a normalized value indicating the co-occurrence likelihood relative to the overall detection of $\text{AP}_i$.


\subsubsection{Dynamic Edge construction}
In previous approaches, the adjacency matrix is built statically, which may lack the flexibility to capture underlying patterns in the data across different training iterations. Also, note that the optimal adjacency matrix construction method can be unpredictable. We propose a third approach that dynamically constructs the adjacency matrix to enhance adaptability. This allows for greater flexibility in capturing complex relationships in the data, potentially leading to better results.

Specifically, we propose a method for dynamic edge construction using K-nearest neighbors (KNN) in batches. This approach leverages the local structure of the data to create edges between similar data points within each training batch.
More specifically, given the total batches $\mathcal{B}$,  for each batch $\boldsymbol b \in \mathcal B$, the dynamic adjacency matrix elements are computed as follows:
\begin{align}
a_{ij}^{\boldsymbol b} &= \begin{cases}
1, & \text{ if } j\in \mathcal N_k^{\boldsymbol b}(i) = top\_k\left (\mathcal D^{\boldsymbol b}, \text{AP}_i  \right ), \\ 
0, & \text{ Otherwise; } 
\end{cases} \\
\mathcal D^{\boldsymbol b} &= \left \{ d_{ij} = \left \| \gamma_{i} - \gamma_{j} \right \|_2\right \}, \left (\text{AP}_i, \text{AP}_j\right) \in \boldsymbol b,
\end{align}
where $\mathcal N_k^{\boldsymbol b}(i)$ 
is the top k nearest neighbors of $\text{AP}_i$, and $\mathcal D^{\boldsymbol b}$ being the pair-wise distance vector in batch $\boldsymbol b$. Further details can be found on this dynamic edge construction in \cite{EdgeConv} where the method was originally proposed.

The utilization of the adjacency matrix in the graph-based model enables the incorporation of valuable spatial information, potentially contributing to the accuracy and robustness of the indoor localization system. Overall, these graph construction approaches leverage the statistical characteristics of the fingerprint database either statically or dynamically, aiming to enhance the model's learning capabilities.

\subsection{Graph Convolution Network}

Building upon the defined indoor localization graph structure, this section explores how graph convolutional networks (GCNs) can be leveraged to improve location estimation accuracy. GCNs exploit the spatial relationships between APs within the graph to achieve this enhancement.

In the context of indoor localization using a GCN, for a given reference position, we define a feature matrix \( \mathbf{X} \in \mathbb{R}^{M \times F} \), where  \( F \) represents the number of features associated with each AP.
Each row of the matrix \( \mathbf{X} \) corresponds to an AP.
%
The features encompass the RSSI and IMU measurements, that is, $F=1+d$, where $d$ is the dimensionality of the IMU sensor data and `$1$' corresponds to the captured RSSI value. For each reference point, a matrix \( \mathbf{X} \) is constructed based on the captured APs at the location and the corresponding IMU data. Thus, the feature of each node (AP) includes the RSSI captured by the device from that AP and the IMU sensing data.
This matrix serves as the input to the GCN, enabling the model to learn spatial dependencies and relationships among the APs through graph convolution. By leveraging the provided features, the GCN facilitates accurate indoor localization. Specifically, \( \mathbf{X} \) is converted into a graph structure, where the adjacency matrix is built using the RSSI data in \( \mathbf{X} \), as described previously. The goal is to extract relevant combined features for the localization task tailored to the network structure.

The GCN layer is a fundamental component that applies a convolution operation on the graph, allowing the model to learn node representations incorporating information from neighboring nodes. The output of the GCN layer, denoted as \( \mathbf{H} \in \mathbb{R}^{M \times C} \), is calculated as follows:
\begin{equation}
    \mathbf{H} = \sigma \left ( \hat{\mathbf{D}}^{-\frac{1}{2}} \hat{\mathbf{A}} \hat{\mathbf{D}}^{-\frac{1}{2}} \mathbf{X} \mathbf{W}\right),
    \label{equ:gcn}
\end{equation}
where \( \hat{\mathbf{A}} = \mathbf{A}_{AP} + \mathbf{I} \) is the adjacency matrix with added self-loops for each node, incorporating both the spatial relationships and self-connections. \( \hat{\mathbf{D}} \) is the diagonal degree matrix of \( \hat{\mathbf{A}} \), where each element \( d_{ii} = \) is the sum of all elements in row \( i \) of \( \hat{\mathbf{A}} \). \( \mathbf{W} \) is a weight matrix of size \( F \times C \), where \( C \) is the number of output features, determining the dimensionality of the node representations. $\sigma (\cdot)$ is the activation function, e.g. the RELU activation.

The convolution operation captures the relationships between nodes in the graph, allowing the model to generate enhanced representations that consider both the graph's inherent structure and the input features. This layer is crucial in leveraging the connectivity information for accurate and context-aware node embeddings.

As shown in Fig. \ref{fig:sysm}, the GCN consists of GC Layers stacked together, enabling the model to capture hierarchical information of the input. The output of each GCN layer serves as the input to the subsequent layer, allowing the network to learn increasingly abstract and informative features from the graph structure.
Going back to \eqref{equ:gcn}, the layer-wise propagation rule can be written as: 
\begin{equation}
    h_{i}^{(l+1)}=\sigma\left(\sum_j \frac{1}{\delta _{i j}} h_{j}^{(l)} W^{(l)}\right),
\end{equation}
where $j$ indexes the neighboring nodes of node $i$ and $\delta _{i j}$ is a normalization constant for the edge $(i,j)$ originating from using the symmetrically normalized adjacency matrix $\hat{\mathbf{D}}^{-\frac{1}{2}} \hat{\mathbf{A}} \hat{\mathbf{D}}^{-\frac{1}{2}}$ in our GCN model. 
The final GCN layer's output provides node embedding for downstream tasks, such as indoor localization. This GCN definition holds when either of the statically built adjacency matrices is employed. However, since our dynamic edge construction adjacency matrix is constructed dynamically, we propose a dynamic graph convolution architecture.

\subsection{Dynamic Edge Convolution}
\label{sec:dec}
Dynamic Edge Convolution (DEC) represents an advancement in graph-based localization systems, aiming to dynamically adapt edge connections within the graph structure to capture evolving spatial relationships among APs. This subsection outlines the formulation and operational principles underlying DEC for localization. The $l^{th}$ edge convolution (EdgeConv) layer is defined as follows:
\begin{equation}
    h_i^{(l+1)}=\bigoplus_{j \in \mathcal{N}(i)}\left[ \psi  \left(\Omega \cdot\left(h_j^{(l)}, h_i^{(l)}\right) + \Phi\cdot h_i^{(l)} \right)\right],
\end{equation}
where $\bigoplus$ the a channel wise symmetric aggregation operation (i.e., $\operatorname{mean}, \operatorname{max}, \operatorname{add}$). $\Omega$ and  $\Phi$ are learnable parameters in the EdgeConv, $\mathcal{N}(i)$ represents the set of neighboring nodes of node $i$. $\psi()$ is the LeakyReLU activation function defined as:
\begin{equation}
    \psi(x) = \operatorname{LeakyRELU}(x) = \begin{cases}
x, & \text{ if } x \geq 0, \\ 
\epsilon \cdot x, & \text{ if } x<0;
\end{cases} 
\end{equation}
where $\epsilon$ controls the angle of the negative slope. In essence, $\psi()$ is an activation function that allows a small, positive gradient when the unit is inactive, helping mitigate the vanishing gradient problem.
Note that $\mathcal{N}(i)$ is computed dynamically for each data batch as explained previously in Section \ref{sec:graphformulation}. Additionally, the representation of an edge $e_{ij}$ can be learned from the node representations using the following equation:
\begin{equation}
    h_{e_{ij}}^{(l+1)}=(h_{i}^{(l+1)} \parallel h_{j}^{(l+1)}).
\end{equation}
This equation shows how the representation of edge $e_{ij}$ at layer $l+1$ is computed by aggregating (i.e, using $\bigoplus$) the representations of nodes $i$ and $j$ at layer $l+1$.

Finally, EdgeConv is the foundation for building a dynamic edge convolutional neural network (DECNN) to learn hierarchical representations of graph-structured localization data. The DECNN consists of multiple layers, each applying the EdgeConv operation to the input graph. The output of each layer is a set of node and edge representations that capture the graph's local and global features.

\subsection{Graph Neural Network based localization }
Following the graph convolution layers (GCN or DEC), multiple fully connected (FC) layers map the learned node embeddings to the final output space. The FC layers operate on the flattened node embeddings and are denoted as:
\begin{align*}
\mathbf{Z}^{(1)} &= \delta(\mathbf{H}_{\text{flatten}} \mathbf{W}_{\text{FC}^{(1)}} + \mathbf{b}_{\text{FC}^{(1)}}) \\
\mathbf{Z}^{(2)} &= \delta(\mathbf{Z}^{(1)} \mathbf{W}_{\text{FC}^{(2)}} + \mathbf{b}_{\text{FC}^{(2)}}) \\
\vdots \\
\mathbf{Z}^{(L)} &= \delta(\mathbf{Z}^{(L-1)} \mathbf{W}_{\text{FC}^{(L)}} + \mathbf{b}_{\text{FC}^{(L)}}),
\end{align*}
where \( \mathbf{H}_{\text{flatten}} \) is the flattened node embeddings matrix $\mathbf{H}$. \( \mathbf{W}_{\text{FC}^{(l)}} \) and \( \mathbf{b}_{\text{FC}^{(l)}} \) are the weight matrix and bias vector of the \(l\)-th FC layer, respectively. \( \delta(\cdot) \) is the activation function applied element-wise. \( \mathbf{Z}^{(l)} \)  is the output of the \(l\)-th FC layer, with $\mathbf{Z}^{(L)} = \mathbf{\hat y}$ the final output (i.e., the predicted location coordinates). 

Let $\boldsymbol \Psi$ represent the FC layers and $g(\cdot)$ the neural network approximator function parameterized by $\boldsymbol \Psi$. The final prediction is given by \mbox{$\mathbf{\hat y} = g(\mathbf{H}_{\text{flatten}};\boldsymbol \Psi)$}. Similarly, we represent by $\boldsymbol \Xi $ the graph convolution layers that parameterize the function $\varphi(\cdot)$ previously defined in \eqref{eq:xfusion}. It follows that \mbox{$\mathbf{H}_{\text{flatten}} = \varphi(\boldsymbol X^{RSSI},\boldsymbol X^{IMU};\boldsymbol \Xi)$}.

The graph convolution layers $\boldsymbol \Xi$ (GCN or DEC layers) and the FC layers $\boldsymbol \Psi$ constitute our GNN, $\boldsymbol{\Theta}$. The graph optimization problem is formulated as optimizing the parameters \( \boldsymbol{\Theta} \) of the GNN to minimize a specific loss function \( \mathcal{L} \). This yields the following optimization problem:
\begin{equation}
\label{eq:gnn}
    \boldsymbol\Theta^* = \underset{\boldsymbol\Theta}{\operatorname{argmin}} \ \mathcal{L}(\mathbf{\hat y},  \mathbf{y}):= \sum_{b\in \mathcal{B}} \sum_{\left (\boldsymbol X,\boldsymbol y  \right )\in b} \mathcal{L}\left ( f\left ( \boldsymbol X;\boldsymbol\Theta \right ), \boldsymbol y \right ),
\end{equation}
where:
\begin{itemize}
    \item \( \boldsymbol\Theta^* \) represents the optimal parameters of the GNN.
    \item \( \boldsymbol X =\left (\boldsymbol X^{RSSI},\boldsymbol X^{IMU} \right) \) is the RSSI and IMU measurement data.
    \item \( f \) is neural network approximator function that parameterize $\boldsymbol\Theta$ and is defined by the following composition: \mbox{\( f\left ( \boldsymbol X; \boldsymbol \Theta \right )= g\left ( \varphi\left ( \boldsymbol X;\boldsymbol \Xi \right ); \boldsymbol \Psi \right ) \) }
    \item \( \mathbf{y} \) is the ground truth, containing the true locations of IoT devices.
\end{itemize}
This optimization process involves adjusting the network's weights and biases to minimize the difference between predicted locations and the ground truth. We employ the well-established stochastic gradient descent (SGD) algorithm, which iteratively updates the parameters to improve the model's efficacy.

%% file: sections/MetaGraphLoc.tex
\section{MetaGraphLoc: Graph meta-learning for Localization}
\label{sec:metagraphloc}
\begin{figure}[!t]
    \centering
    \includegraphics[width=\linewidth]{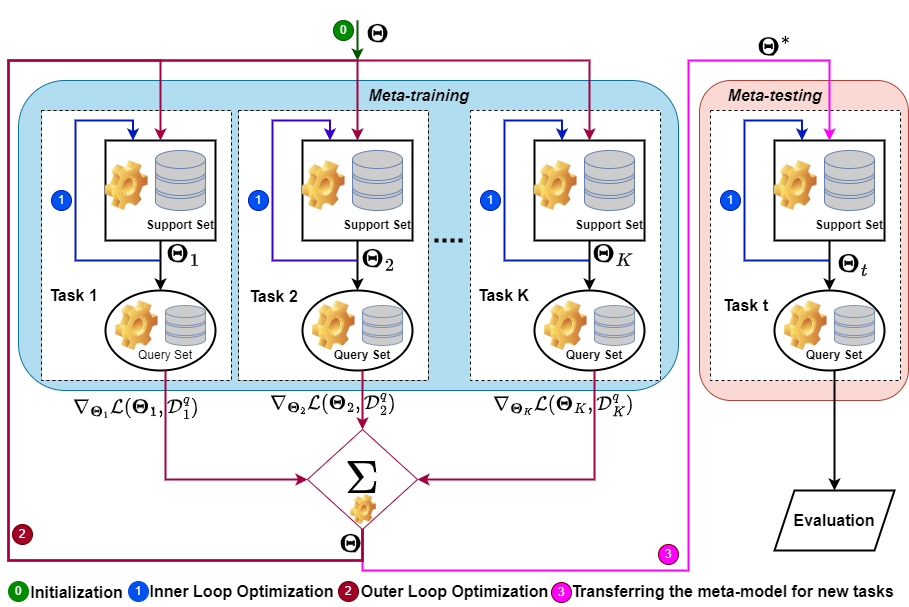}
    \caption{Meta-learning system architecture.}
    \label{fig:meta}
\end{figure}
Previously, we constructed a GNN model that can learn feature representations in indoor environments, facilitating the development of efficient indoor positioning systems. However, traditional approaches typically tailor models to specific indoor environments, accounting for their unique structures, dynamics, wireless conditions, and technologies.
Building upon the limitations of training a separate GNN model for each unique indoor environment, we introduce MetaGraphLoc, a graph-based meta-learning framework for indoor localization. This approach leverages meta-learning to overcome the challenges of limited training data and environment-specific signal variations often encountered in indoor positioning systems. This section explores graph meta-learning within the MetaGraphLoc framework, whose general architecture is depicted in Fig. \ref{fig:meta}.

 \subsection{Graph meta-learning for localization}
 Meta-learning, also known as ``learning to learn,'' aims to train models that can quickly adapt to new tasks or environments with a limited amount of data \cite{maml}. Unlike traditional ML methods, which focus on optimizing a single task, meta-learning algorithms learn from diverse tasks and extract transferable knowledge that enables them to learn new tasks efficiently.
 
 \subsubsection{Task definition}
 In our proposed framework, we defined a task as floor-level localization. 
 In indoor localization, the floor-level localization task involves map sensor measurements, such as WiFi signal strengths or IMU readings, to learn specific x-y coordinates within a designated floor of an indoor environment. Given a defined floor area of interest, the goal is to accurately determine a target's horizontal position (x, y coordinates) based on the available sensor data.
 
 Formally, we define the floor-level localization task by $\mathcal{T}(\mathcal{D},\mathcal{L})$ where:
 \begin{itemize}
     \item $\mathcal{D}=(\mathcal{D}^s, \mathcal{D}^q)$ represents the dataset containing sensor measurements, namely  RSSI and IMU readings. $\mathcal{D}^s$ is the support set used to train the model in the inner loop (this corresponds to the training set in convention ML). $\mathcal{D}^q$ is the query set used to evaluate the model after the inner loop, and the evaluation loss is used to update the model in the outer loop.
     \item $\mathcal{L}$ denotes the loss associated with the learning model, which measures the discrepancy between the predicted coordinates and the ground truth.
 \end{itemize}
 Additionally, we designate by $\mathbf{\Theta}$ the GNN associated with the localization task. 

\subsubsection{Meta-learning objective and training}
Meta-learning aims to train a meta-model with training tasks that can adapt quickly to new localization tasks (test tasks).
Meta-learning enables the model to adapt its parameters to new tasks quickly, provided that the training and testing tasks share sufficient similarity. We consider the floor-level localization task $\mathcal{T}$ as defined earlier, where the goal is to predict the floor-level coordinates based on RSSI and IMU.

Assume that we have access to the meta-training dataset $\mathcal{T}_{train}$ containing $K$ localization tasks, i.e., $$\mathcal{T}_{train} = \left\{  \mathcal{T}_k=(\mathcal{D}_k, \mathcal{L}_k) \right\}_{k=1,2,\cdots, K}.$$

The meta-learning objective is a bi-level optimization problem  defined as
\begin{equation}
\label{equ:objective}
    \min _\mathbf\Theta J(\mathbf\Theta):= \sum_{k=1}^K \mathcal{L}\left(\underbrace{\mathbf\Theta-\mu_k \nabla_\mathbf\Theta \mathcal{L}_k\left(\mathbf\Theta, \mathcal{D}_k^s\right)}_{\mathbf\Theta_k}, \mathcal{D}_k^q\right),
\end{equation}
where $\mu_k$ is the inner learning rate of task $\mathcal{T}_k$, $\mathcal{L}$ the global loss function associated with the meta-learner, and $\nabla$ is the gradient operator. 
Note that $\mathbf\Theta_k$ is the locally updated model resulting from the inner loop optimization (first level of optimization). This update is done iteratively using SGD. Indeed, for all localization tasks in training tasks, we consider $N$ as the total number of SGD for the inner loop update. That is,
\begin{equation}
    \begin{aligned} 
        \mathbf\Theta_k^{n+1} &= \mathbf\Theta_k^{n}-\mu_k \nabla_{\mathbf{\Theta}_k^{n}} \mathcal{L}_k\left(\mathbf\Theta_k^{n}, \mathcal{D}_k^s\right)\\
        & \text{ for } \ \ n = 1,2,\cdots,N-1,
        \end{aligned}
        \label{eq:innerloop}
\end{equation}
where $ \mathbf\Theta_k^{0} = \mathbf\Theta \text{ and } \mathbf\Theta_k^{N} = \mathbf\Theta_k$. 

 After the model is updated for every localization task, \eqref{equ:objective} is also optimized through SGD. This is the second optimization level, known as the outer loop optimization. Specifically, at each iteration, the meta-model is updated as follows.
\begin{equation}
    \begin{aligned} 
        \mathbf\Theta(i+1) &= \mathbf\Theta(i)-\frac{\eta}{K}\sum_{k=1}^K  \nabla_{\mathbf{\Theta}(i)} \mathcal{L}_k\left(\mathbf\Theta_k, \mathcal{D}_k^q\right)\\
        & \text{ for } \ \  i = 1,2,\cdots,I-1
        \end{aligned},
\end{equation}
where $\eta$ is the outer learning rate and $I$ is the total number of iterations.
Additionally, we account for a task distribution $\rho(\mathcal{T})$ to capture the heterogeneity of the training tasks and adjust the update rule accordingly. More specifically, each task would contribute to the global update according to its abundance in data availability. Thus, the global model is updated as follows:
\begin{equation}
    \begin{aligned} 
        \mathbf\Theta(i+1) &= \mathbf\Theta(i)-\eta\sum_{k=1}^K  \rho_k\nabla_{\mathbf{\Theta}(i)} \mathcal{L}_k\left(\mathbf\Theta_k, \mathcal{D}_k^q\right),\\
        & \rho_k = \rho\left ( \mathcal{T}_k \right )= \frac{\left | \mathcal{D}_k^s \right |}{\sum_k \mathcal{D}_k^s }.
        \end{aligned}
        \label{eq:outerloop}
\end{equation}
This outer-loop optimization aims to obtain a reasonable approximation for the optimal $\mathbf\Theta^* $ after $I$ iterations, such that
\begin{equation}
    \mathbf{\Theta}\left ( I \right ) \approx \mathbf{\Theta}^* = \underset{\mathbf{\Theta}}{\operatorname{argmin}} \ J\left ( \mathbf{\Theta} \right ).
\end{equation}
The overall process can be summarized by Algorithm \ref{alg:meta-training}.

\SetKwComment{Comment}{/* }{ */}
\RestyleAlgo{ruled}
\begin{algorithm}[!t]
\SetAlgoLined
  \caption{MetaGraphLoc training phase.}
  \label{alg:meta-training}
  \SetKwInOut{Input}{Inputs}
  \SetKwInOut{Output}{Outputs}
  \Comment{Input: Set of localization tasks (RSSI datasets) to train the meta-model}
  \Input{ \\ - $\mathcal{T}_{train} = \{\mathcal{T}_k = \{\mathcal{D}_k, \mathcal{L}_k\}\}$\;
  - $I$: Number of iterations\;
  - $\mu$: Learning rate for local training (inner)\;
  - $\eta$: Learning rate for global updates (outer)\;
  }
  \Comment{Output: Trained meta-model}
  \Output{$\{\boldsymbol{\Theta}^*\}$}
  
  \SetKwProg{ServerInit}{Initialization}{}{}
  \ServerInit{($\eta$, $\mu, N, I$)}{
   $ i \gets 0$ , initialize $\boldsymbol{\Theta}(0) $ randomly\;
  Task-specific models' configuration: $\boldsymbol{\Theta}_k \gets \boldsymbol{\Theta}(0)$\;
  $K \gets \left | \mathcal{T}_{train} \right |$ and  $ \mu_k \gets \mu ,\text{ for } k = 1, 2, \cdots, K $\;
  
  }
  \SetKwProg{TrainingLoop}{Main Loop}{}{}
  \TrainingLoop{()}{
      \For{$i \gets 1$ \KwTo $I$ }{
      
      \SetKwProg{ClientTraining}{Task-specific Training}{}{}
           \ClientTraining{$(\boldsymbol{\Theta(i-1)}, \mu_k,\mathcal{T}_{train}$)}{
           \ForEach{$\mathcal{T}_k$ $\in \mathcal{T}_{train}$}{
           Task-specific models' synchronization: 
            $ n \gets 0$,  $\boldsymbol{\Theta}_\kappa^{n} \gets \boldsymbol{\Theta}(i-1)$\;
    
            \For{$n \gets 1$ \KwTo $N$}{
            Train the task model using \eqref{eq:innerloop}\;
            }
            Get trained model $\boldsymbol{\Theta}_k = \boldsymbol{\Theta}_k^N$\;
      }
      }
    \SetKwProg{GlobalUpdate}{Global Update}{}{}
          \GlobalUpdate{($\eta$, $\{\boldsymbol{\Theta}_k\}$)}{
           Update the global model 
           $\mathbf\Theta(i) = \mathbf\Theta(i-1)-\frac{\eta}{K}\sum_{k=1}^K  \nabla_{\mathbf{\Theta}(i-1)} \mathcal{L}_k\left(\mathbf\Theta_k, \mathcal{D}_k^s\right)$
    }
      
  }
  }
  \end{algorithm}

\subsubsection{Meta-testing}
After the meta-training, the trained meta-model $\mathbf{\Theta}^*$ is ready to be deployed and adapted to new indoor environments.
To assess the effectiveness of meta-learning, we deploy the pre-trained meta-model for new localization tasks, which constitute the meta-testing dataset. We assume $T$ available localization tasks in this dataset $\mathcal{T}_{test}$ such that
$$\mathcal{T}_{test} = \left\{  \mathcal{T}_t=(\mathcal{D}_t, \mathcal{L}_t) \right\}_{t=1,2,\cdots, T}.$$
Then for each task $\mathcal{T}_t \in \mathcal{T}_{test}$, the
task-speciﬁc model is initialized with  $\mathbf{\Theta}^*$ and $J$ steps of gradient descent are performed on small support set $\mathcal{D}_t^s$, to obtain task-specific adapted parameters $\mathbf{\Theta}^J_t$ that can be
expressed as:
\begin{equation}
    \boldsymbol{\Theta}_t^J = \boldsymbol{\Theta}^* - \mu_t \left[ \nabla_{\boldsymbol{\Theta}^*} \mathcal{L}_t \left( \boldsymbol\Theta^* ; \mathcal D_t^s \right) + \sum_{j=1}^{J-1} \nabla_{\boldsymbol{\Theta}_t^j} \mathcal{L}_t \left( \boldsymbol\Theta_t^j ; D_t^s \right) \right].
\end{equation}
The optimal parameters $\boldsymbol{\Theta}_t^*$ should satisfy the formal optimization problem. That is \mbox{$ \boldsymbol{\Theta}_t^* = \underset{\mathbf{\Theta_t}}{\operatorname{argmin}} \mathcal{L}_t\left(\mathbf\Theta_t, \mathcal{D}_t^s\right).$}

The objective is to drive $\boldsymbol{\Theta}_t^J$  as close as possible to $\boldsymbol{\Theta}_t^*$. More specifically, it is about finding the minimal $J$ that minimizes the residual function $\mathcal Q_t(J)$ for task $t$, and defined over all the test tasks by
\begin{equation}
    \mathcal Q(\boldsymbol J) =  \sum_{t=1}^{T}\underbrace{\left \| \mathcal{L}_t \left( \boldsymbol\Theta_l^{J_t} ; D_t^q \right) - \mathcal{L}_t \left( \boldsymbol\Theta_t^* ; D_t^q \right)\right \|^2}_{\mathcal Q_t(J)}.
\end{equation}
To obtain the minimal values of $\boldsymbol J$, we define an accuracy-threshold $\epsilon$ and  $\left\{\mathcal Q_t(J)\right\}$ are said to be $\epsilon-$accurate, where
\begin{equation}
    \boldsymbol J = \Bigl[ J=\min \left \{ j\mid  \mathcal Q_t(j)<\epsilon  \right \}, \  t=1,2,\cdots,T\Bigr].
\end{equation}
Smaller values of $\boldsymbol J$ (i.e., few SGD steps) imply fewer data collection and calibration efforts, a key motivation for meta-learning. 

\SetKwComment{Comment}{/* }{ */}
\RestyleAlgo{ruled}
\begin{algorithm}[!t]
\SetAlgoLined
  \caption{MetaGraphLoc testing phase.}
  \label{alg:meta-testing}
  \SetKwInOut{Input}{Inputs}
  \SetKwInOut{Output}{Outputs}
  \Comment{Input: Set of test tasks}
  \Input{ \\ - $\mathcal{T}_{test} = \{\mathcal{T}_t = \{\mathcal{D}_t, \mathcal{L}_t\}\}$\;
  - $\mu_t$: Learning rate for local training (inner)\;
  - $\{\boldsymbol{\Theta}^*\}$: Pretrained meta-model
  }
  \Comment{Output: Trained local model}
  \Output{$\{\boldsymbol{\Theta}^J_t\}$}

 \SetKwProg{ClientTraining}{Task-specific Training}{}{}
           \ClientTraining{$( \mu_t,\mathcal{T}_{test}, \epsilon$)}{
          
           \ForEach{$\mathcal{T}_t$ $\in \mathcal{T}_{test}$}{
              \SetKwProg{ServerInit}{Initialization}{}{}
              \ServerInit{($\mu_t, \boldsymbol{\Theta}^*$)}{
               
               $ j \gets 0$\;
              knowledge transfer: $\boldsymbol{\Theta}_t^j \gets \boldsymbol{\Theta}^*$\; 
              }
            \While{$ \mathcal Q_t(j)<\epsilon $}{
            $\mathbf\Theta_t^{j+1} = \mathbf\Theta_t^{j}-\mu_t \nabla_{\mathbf{\Theta}_t^{j}} \mathcal{L}_t\left(\mathbf\Theta_t^{j}, \mathcal{D}_t^s\right)$\;
            $j \gets j+1$\;
            }
            Get trained model: $\boldsymbol{\Theta}_t^J = \boldsymbol{\Theta}_t^j$ (i.e., $J \gets j$)\;
      }
      }
  \end{algorithm}

\subsection{Environment adaptation}
Traditional meta-learning assumes a consistent distribution of task data, a presumption impractical within indoor localization contexts characterized by diverse geometries and, notably, varying numbers of APs. These differences result in distinct signal spaces, fundamentally shaping the distributions of fingerprint datasets. To address this challenge, we propose aligning the signal spaces of training tasks through dimensionality reduction techniques. Specifically, we aim to transform the input spaces of each indoor environment into a shared latent representation.

To this end, MetaGraphLoc incorporates environment adaptation using principal component analysis (PCA) for dimensionality reduction. PCA transforms the input signal spaces of each indoor environment (tasks) into a common latent representation (corresponding to the most relevant APs), facilitating better adaptation during meta-testing. This transformation involves the following steps:
\subsubsection{Normalizing the data} $\boldsymbol{X}^\prime  = \frac{\boldsymbol X-\bar{\boldsymbol X}}{\boldsymbol\Phi}$.\\
We subtract the mean of each feature, $\bar{\boldsymbol X}$, from the dataset to centre it around zero and scale using the standard deviation $\boldsymbol\Phi$.
\subsubsection{Computing the Covariance Matrix} $\boldsymbol\Delta  = \frac{1}{n}\left ( \boldsymbol{X}^\prime  \right )^T\boldsymbol{X}^\prime $.\\
We calculate the covariance matrix $\boldsymbol\Delta$ of the normalized data $\bar{\boldsymbol X}$, which captures the relationships between different features.
\subsubsection{Eigen decomposition} $\boldsymbol\Delta\boldsymbol V  = \boldsymbol\Lambda \boldsymbol V$.\\
We compute the eigenvalues $\boldsymbol\Lambda$ and eigenvectors $\boldsymbol V$ of the covariance matrix. The eigenvectors represent the principal components, and the eigenvalues indicate the variance each component measures.
\subsubsection{Transforming the Data} $\boldsymbol{X}_{pca} = \boldsymbol{X}^\prime\boldsymbol V_m$.\\
The top $m$ eigenvectors, $\boldsymbol V_m$, are selected; those eigenvectors correspond to the highest eigenvalues and retain the most significant information while reducing dimensionality. Subsequently, we project the original normalized data $\boldsymbol{X}^\prime$ onto the selected principal components to obtain the new, reduced-dimensional representation $\boldsymbol{X}_{pca}$.

MetaGraphLoc addresses the challenges arising from different signal space dimensions across indoor environments by transforming the input signal spaces into a common latent representation using PCA. This allows the meta-model to learn transferable knowledge applicable to unseen environments with potentially varying numbers of APs, facilitating more effective adaptation during meta-testing.


%% file: sections/performance-eval.tex
\section{Performance evaluation and discussion}

\label{sec:perfeval}
\subsection{Dataset description}
\begin{figure}[!t]
    \centering
    \includegraphics[width=0.8\linewidth]{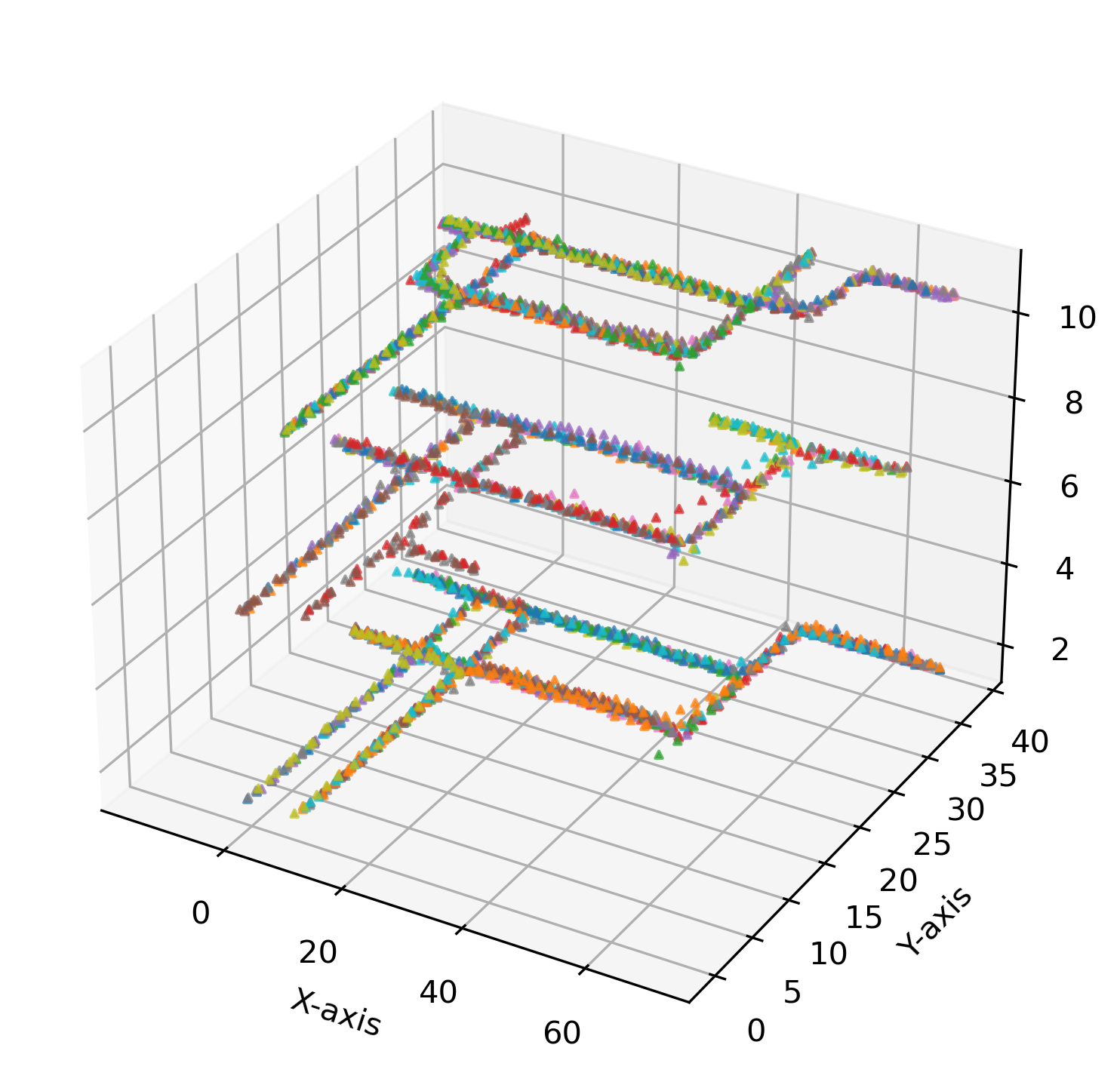}
    \caption{IMUWiFine sampled Dataset visualization. Different colors indicate different trajectories with sampled data points (sampling step=100).}
    \label{fig:dataviz}
\end{figure}

To evaluate the performance of our proposed solution, we adopt the IMUWiFine experimental dataset \cite{e2eseq}. The IMUWiFine dataset is a comprehensive collection of indoor localization data, integrating IMU and WiFi RSSI measurements. It comprises 120 trajectories recorded on various days and times across 3 floors of a 9,000 $m^2$ building, using a mobile phone equipped with IMU and WiFi capabilities. Fig \ref{fig:dataviz} provides a snapshot of the data distribution across different floors and trajectories. These trajectories capture diverse movement patterns, speeds, and environmental conditions, offering fine-grained spatio-temporal information. Each sample is timestamped using the phone's operating system, guaranteeing accurate temporal alignment.


It is worth noting that the high amount of data collected results in a large dataset, potentially hindering its generalization in other indoor environments (i.e., overfitting on this specific environment). Therfore, we employ a down-sampling strategy to address this and enable our solution to function effectively even with less data. Instead of using the full trajectory data, we sampled data points from each trajectory as illustrated in Fig. \ref{fig:dataviz} with a sampling step of 100. This reduction in data points ensures that the model's performance is not solely attributed to abundant data availability.

The dataset is divided into three sets: a training set consisting of 60 trajectories with 3.1 million samples, a validation set with 30 trajectories and 1.2 million samples, and a test set with 30 trajectories and 1.1 million samples. It is available in two formats: raw data and pre-processed data. In the pre-processed data, the number of APs is 379, making the RSSI fingerprint vectors 379-dimensional. The IMU data consists of 9 measurements from three components—accelerometer, gyroscope, and magnetometer—along three dimensions. Consequently, each sample comprises 379 RSSI measurements, 9 IMU measurements, and the corresponding 3D location coordinates (x, y, z), where z indicates the floor level. For our experiments, we utilize subsets (i.e., data points sampling) of two main sets: the training set with 307,185 samples and the validation set with 229,974 samples.

\subsection{RSSI and IMU data fusion}
\begin{figure}[!t]
    \centering
    \includegraphics[scale=0.8]{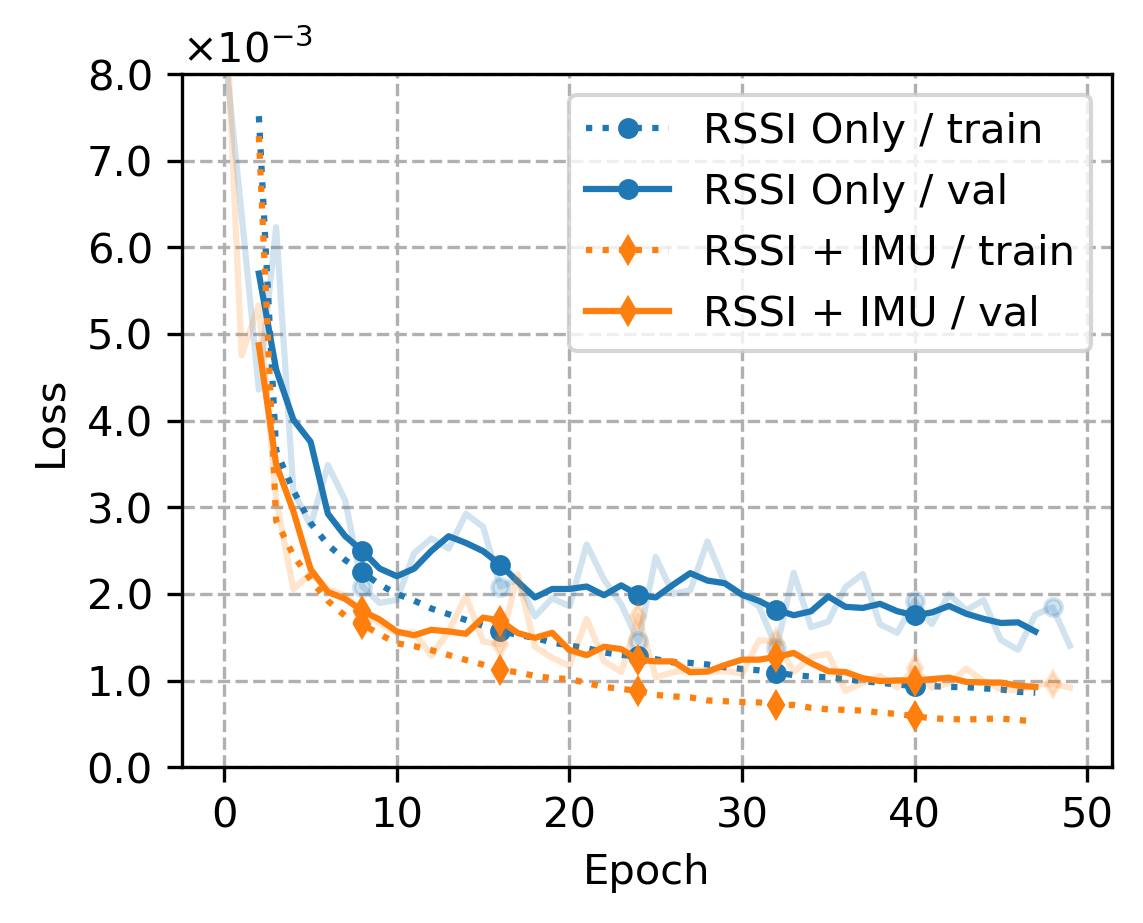}
    \caption{Evaluation of the data fusion. method}
    \label{fig:fusion}
\end{figure}
To assess the value of data fusion in MetaGraphLoc, we evaluate a proposed GNN architecture, DEC, on two configurations: using WiFi RSSI measurements alone and using both RSSI and IMU data combined. The DEC-based GNN model consists of two edge Convolution layers with 128 units each. We employ the ADAM optimizer with a learning rate of 0.0005 and use the mean squared error (MSE) as the loss function. Training is conducted for 50 epochs with a batch size of 8. All the parameters used in the experiments are summarized in TABLE \ref{tab:config}.

\begin{table}[!t]
\label{tab:config}
\caption{Experiment parameters configuration.}
\begin{tabular}{lll}
\toprule
\textbf{Parameter} & \textbf{Description}                 & \textbf{Value} \\ \midrule \midrule
m                  & Meta input dimension                 & 120            \\ \hline
$K_{neigh}$        & KNN-graph in DEC                     & 15             \\ \hline
Opt        & Optimizer                     & ADAM             \\ \hline
$\eta$             & Outer learning rate                  & 0.001          \\ \hline
$\mu$              & inner learning rate                  & 0.0005         \\ \hline
s                  & Trajectory data points sampling step & 100            \\ \hline
L                  & Number of  hidden layers             & 2              \\ \hline
h                  & Hidden layer size                    & 128            \\ \hline
N                  & Number of local SGD steps            & 5              \\ \hline
I                  & Number of global iterations          & 1500           \\ \hline
b                  & batch size                           & 8              \\ \hline
K                  & Total number of  training tasks      & 2              \\ \bottomrule
\end{tabular}
\label{tab:config}
\end{table}

Fig. \ref{fig:fusion} presents the training performance for both configurations. It demonstrates that fusing RSSI and IMU data leads to faster convergence and superior localization accuracy. Additionally, after 50 epochs (epoch 0 to epoch 49), the evaluation shows that the model achieves an accuracy down to 2.43 meters with data fusion compared to 2.89 meters using only RSSI. This translates to a significant 16\% reduction in localization error. These results highlight the effectiveness of data fusion for enhancing indoor localization performance. Therefore, we adopt the fusion of RSSI and IMU measurements for the rest of the experiments in MetaGraphLoc.

 %
  %

\subsection{GNN models evaluation}

\begin{table}[!t]
\caption{Errors (in meters) for different thresholds for Adjacency matrix construction: GCN-prob (Spatial relationship-based), GCN-corr (Correlation-based), DEC (Dynamic edge convolution).}
 \centering
\begin{tabular}{llllll}
\toprule
\textit{Threshold} &\textit{ 0.1}  & \textit{0.2}           & \textit{0.3}  & \textit{0.4}  & \textit{0.5 } \\ \midrule \midrule

GCN-prob  & 4.39 & \textbf{4.28} & 4.35 & 4.36 & 4.38 \\ \hline
GCN-corr  & 4.15 & \textbf{4.12} & 4.14 & 4.20 & 4.13 \\ 
\bottomrule
 &  &  &  &  & \\ 
\textit{$K_{neigh}$}  & \textit{5} & \textit{10} & \textit{15} & \textit{20} & \textit{25} \\ \midrule \midrule
DEC  & 2.98 & 2.82   & \textbf{2.60}  & 2.63  & 2.66  \\ \bottomrule
\end{tabular}
\label{tab:thr}
\end{table}

\begin{figure}[!t]
    \centering
    \includegraphics[scale=0.8]{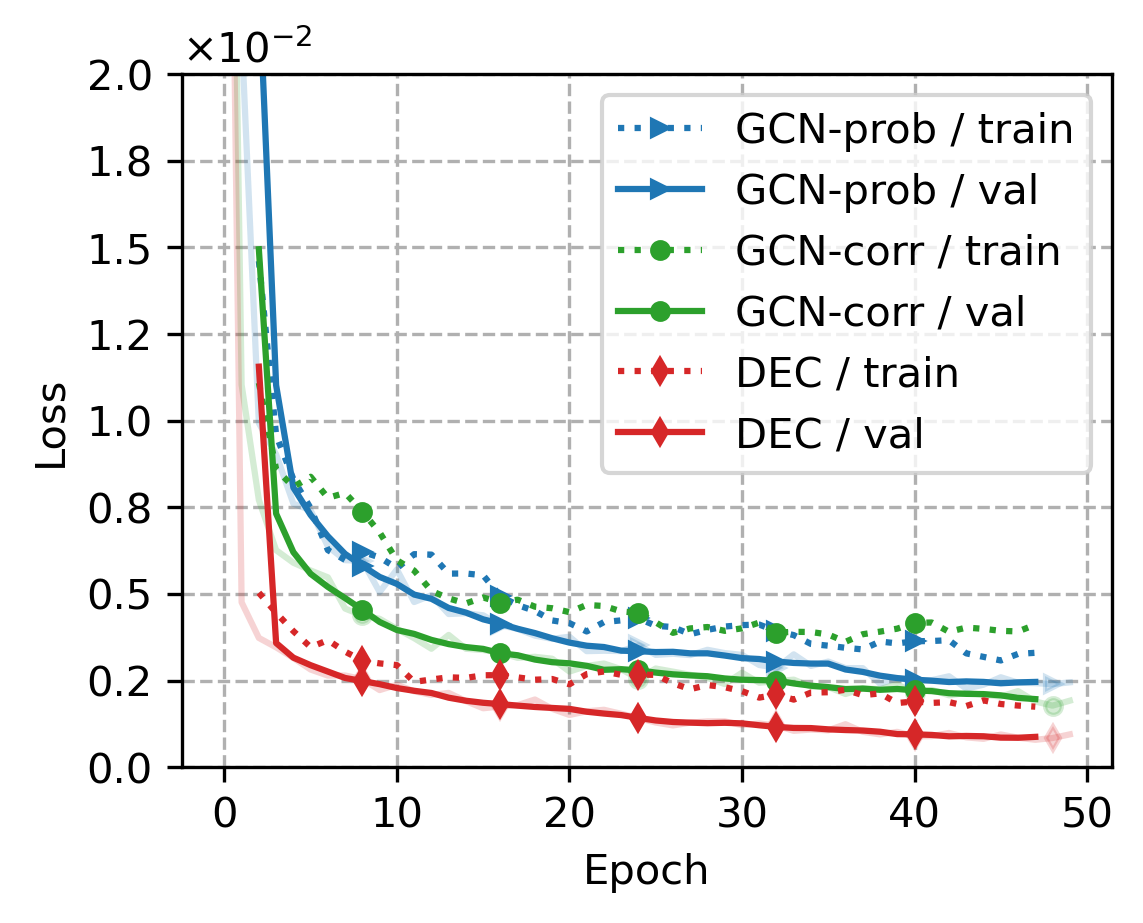}
    \caption{Loss function for different adjacency matrix construction methods.}
    \label{fig:adjM}
\end{figure}

\begin{figure}[!t]
    \centering
    \includegraphics[scale=0.8]{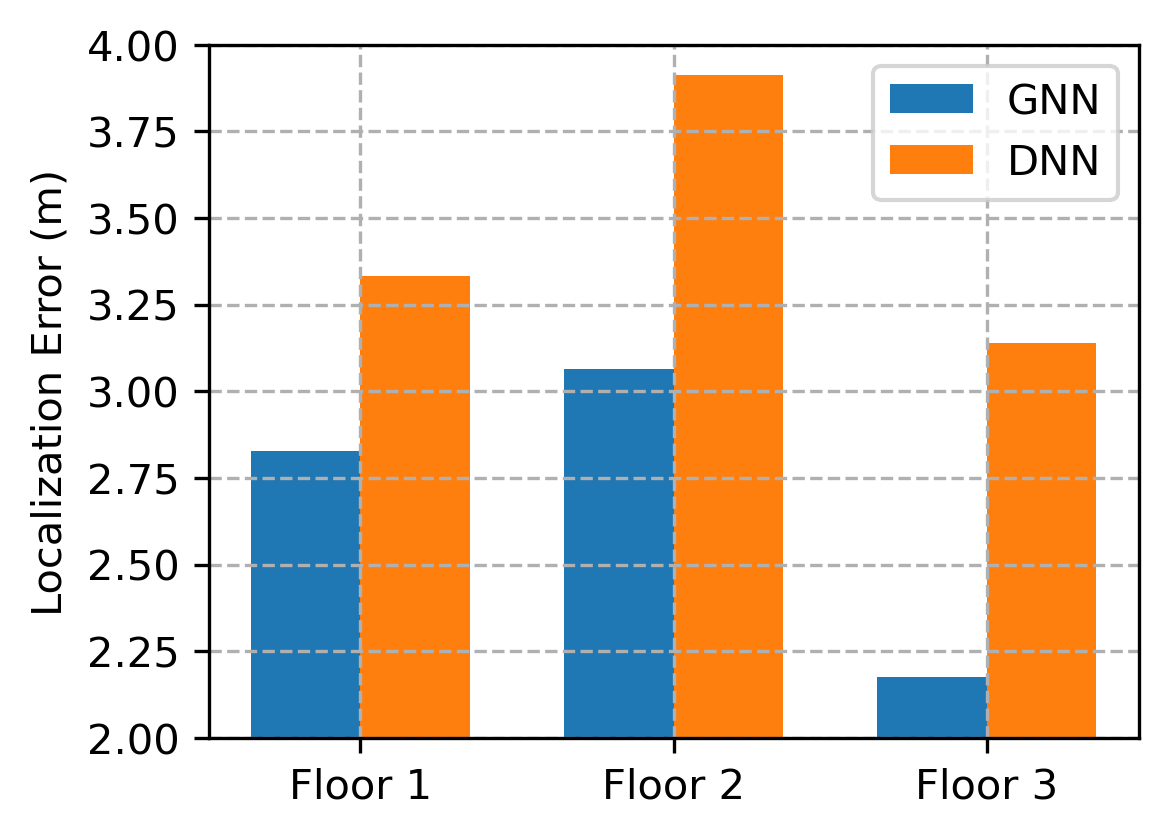}
    \caption{Graph-based approach vs. DNN approach.}
    \label{fig:versus}
\end{figure}

This section first evaluates the performance of different adjacency matrix construction methods for our GNN model. We focus on the two proposed static methods: correlation-based and spatial relationship-based. For both methods, we explore a range of thresholds (0.1 to 0.5) to determine whether there is a relationship between two APs based on statistical measures, aiming to identify the optimal value that maximizes performance. Additionally, we evaluate the DEC method, which dynamically builds the adjacency matrix using a KNN graph. We fine-tune this approach by varying the parameter $K_{neigh}$, which determines the number of neighboring nodes considered for each node.

As shown in TABLE \ref{tab:thr}, the correlation-based approach emerges as the best static method, achieving its peak performance with a threshold of $0.2$. However, it falls short of the dynamic DEC method ($K_{neigh}=15$) in terms of accuracy. Furthermore, the results in Fig. \ref{fig:adjM} confirm that the DEC method achieves lower loss and thus, superior accuracy for indoor localization tasks compared to other methods with their optimized thresholds. 
Consequently, we adopt DEC as the adjacency matrix construction method for the remainder of our experiments.

Building on the superiority of the DEC adjacency matrix construction method, we now evaluate the effectiveness of the proposed GNN model compared to a standard DNN approach. 
The results depicted in Fig. \ref{fig:versus}, demonstrate that GNN consistently outperforms the DNN across all floors. More specifically, the GNN achieves $15.31\%$, $21.74\%$, and $30.89\%$ reductions in localization errors compared to DNN, on floor 1, 2, and 3 respectively. This finding reinforces the advantage of DEC aided graph-based approaches for indoor localization tasks compared to traditional DNNs.

\begin{figure*}[!t]
    \centering
    \begin{subfigure}{0.33\textwidth}
        \includegraphics[width=\linewidth]{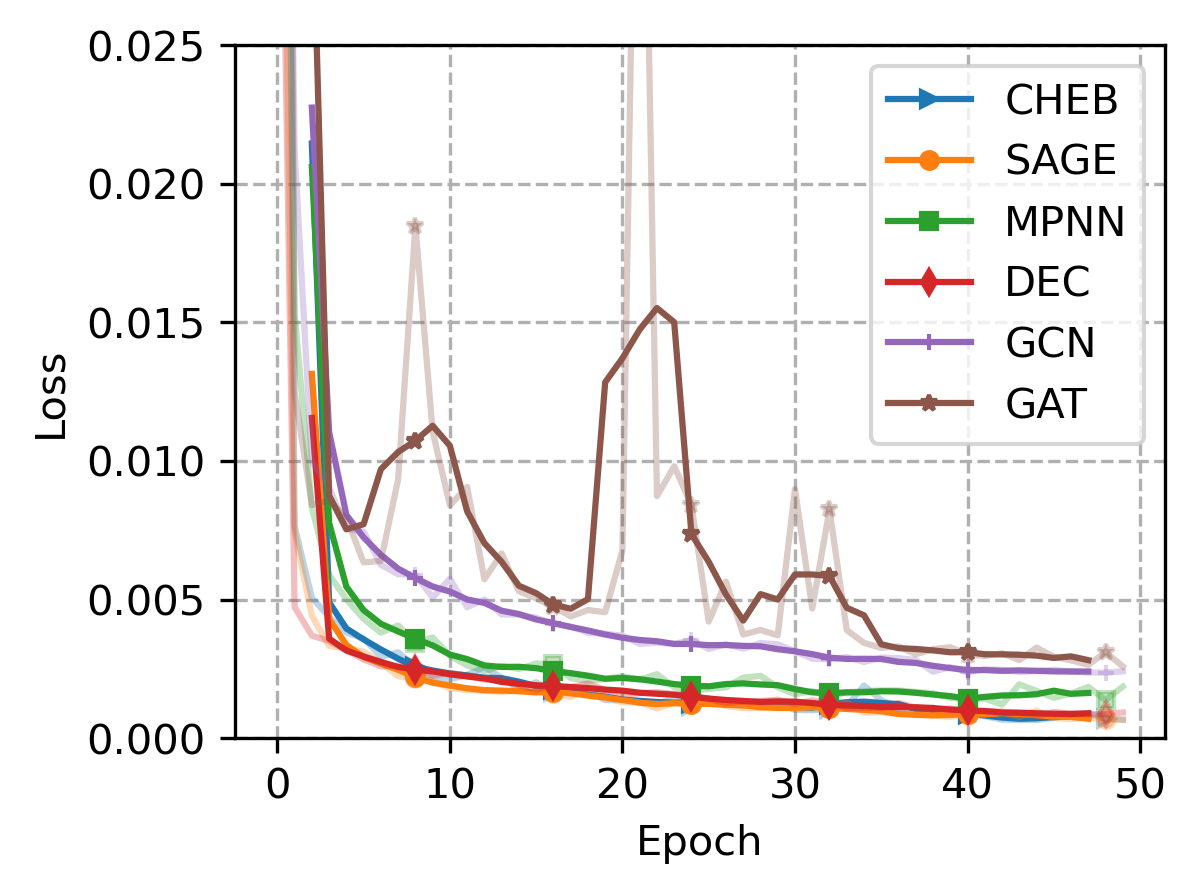}
        \caption{Training performance}
        \label{fig:loss}
    \end{subfigure}
    \hfill
    \begin{subfigure}{0.33\textwidth}
        \includegraphics[width=\linewidth]{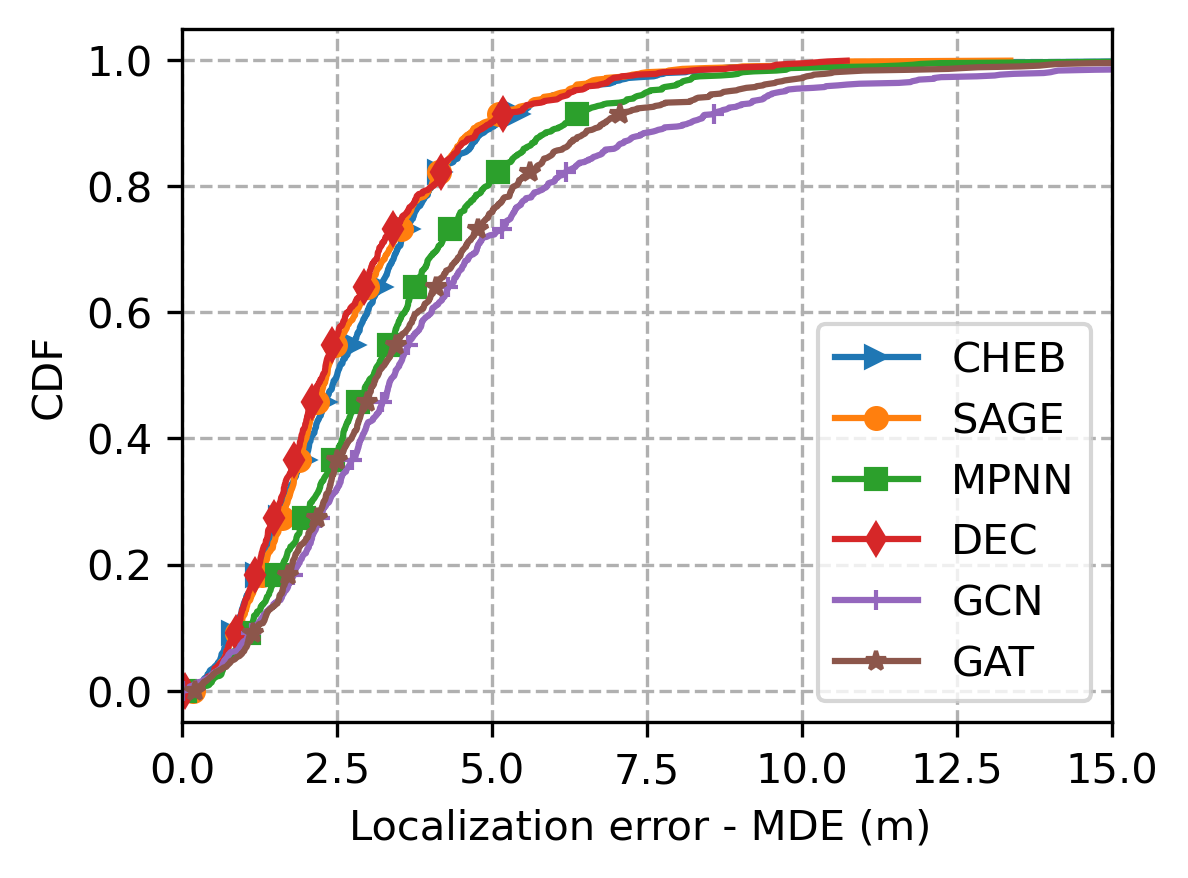}
        \caption{CDF Plots}
        \label{fig:cdf}
    \end{subfigure}
    \hfill
    \begin{subfigure}{0.32\textwidth}
        \includegraphics[width=\linewidth]{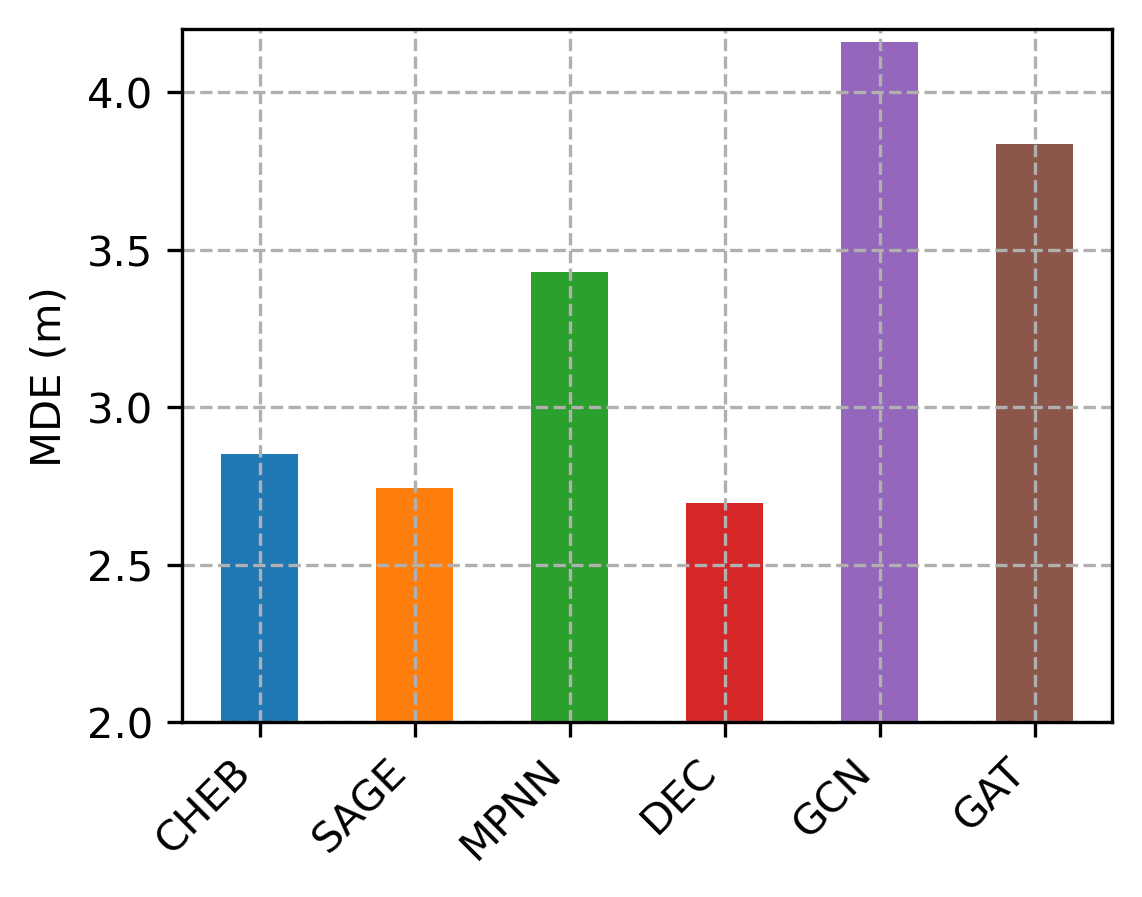}
        \caption{MDE comparision}
        \label{fig:mde}
    \end{subfigure}
    \caption{Comparison of different GNN model architectures for indoor localization.}
    \label{fig:benchmk}
\end{figure*}

We further evaluate the performance of the proposed GNNs (GCN and DEC) by comparing them to established baseline GNN architectures. These include the basic Message Passing Network (MPNN) \cite{MPNN}, Graph Attention Network (GAT) \cite{GAT}, ChebNet, and GraphSAGE \cite{GraphSAGE}. We compare these models based on their convergence behavior, visualizing the training loss function across epochs in Fig. \ref{fig:loss}. This plot highlights the convergence behavior of each model, indicating how efficiently they learn from the data. DEC consistently shows the lowest loss throughout the epochs, suggesting that it learns more efficiently compared to the other models.

To comprehensively evaluate localization accuracy, we generated cumulative distribution functions (CDFs) for each GNN model in Fig. \ref{fig:cdf}. These CDFs depict the distribution of localization errors, providing insights into the models' ability to predict locations across the dataset accurately. Compared to the other methods, DEC exhibits a steeper rising curve, indicating that a higher percentage of its predictions fall within a smaller error range. This suggests that this model offers better localization accuracy with higher probability. Finally, we quantify the average localization error for each GNN model using the mean distance error (MDE). Fig. \ref{fig:mde} presents a bar plot of the MDE, offering a clear comparison of their overall localization accuracy. As expected, DEC achieves the lowest MDE ($\approx 2.6$ meters), further confirming its superior performance in accurately predicting locations.

\subsection{Meta-learning: MetaGraphLoc Evaluation }
\begin{figure*}[!t]
    \centering
    \begin{subfigure}{0.33\textwidth}
        \includegraphics[width=\linewidth]{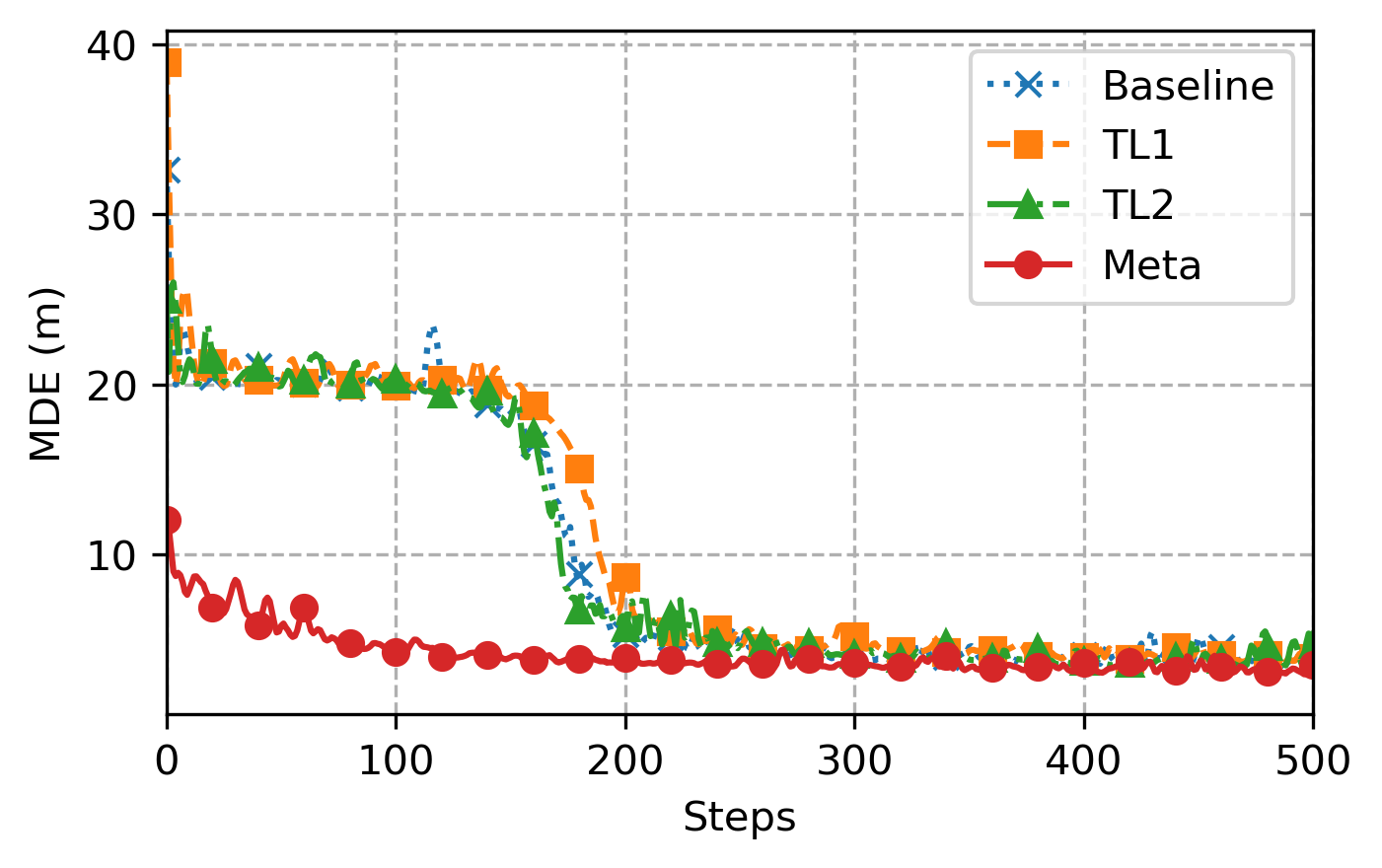}
        \caption{Meta-testing on floor 3}
        \label{fig:meta3}
    \end{subfigure}
    \hfill
    \begin{subfigure}{0.33\textwidth}
        \includegraphics[width=\linewidth]{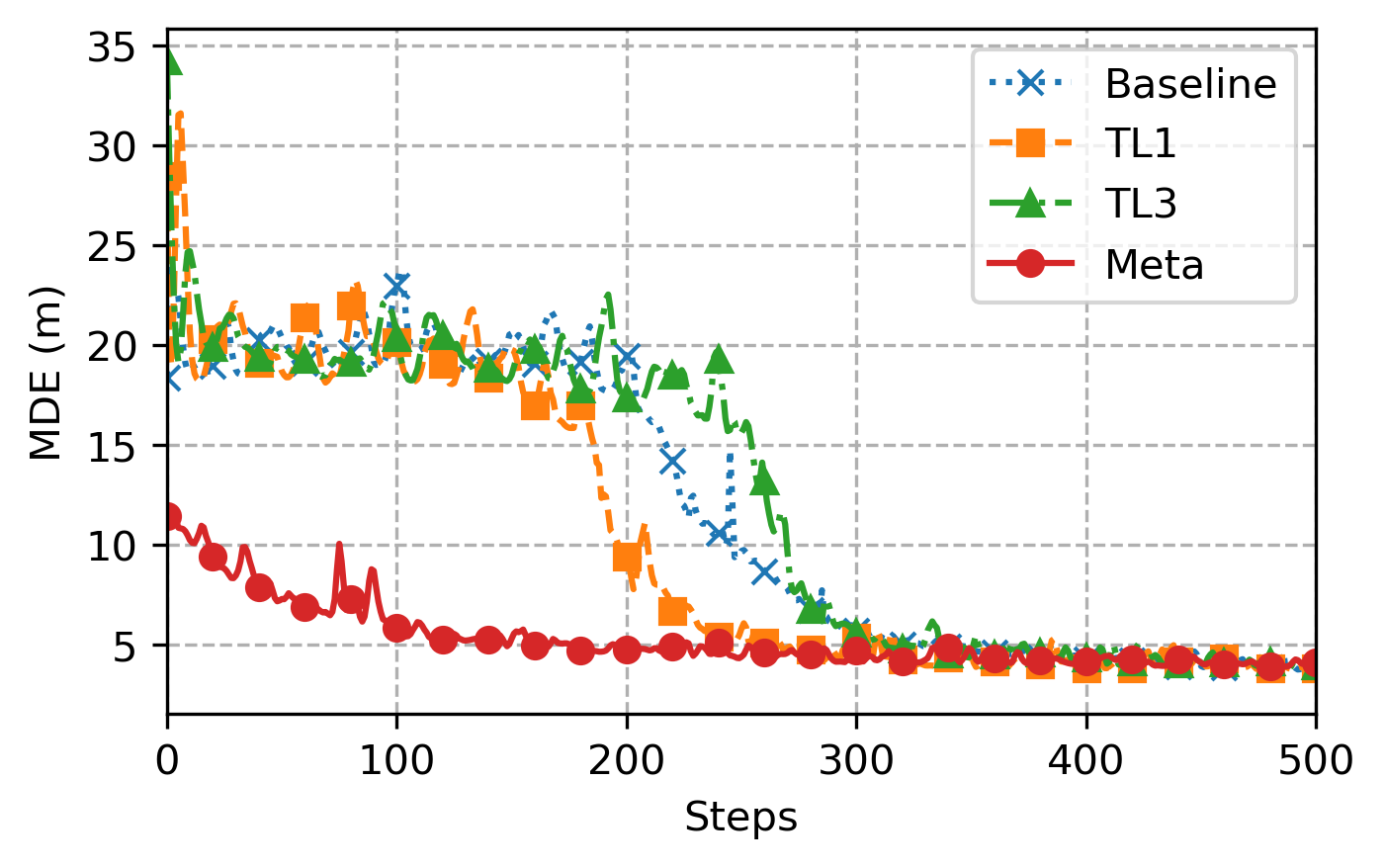}
        \caption{Meta-testing on floor 2}
        \label{fig:meta2}
    \end{subfigure}
    \hfill
    \begin{subfigure}{0.32\textwidth}
        \includegraphics[width=\linewidth]{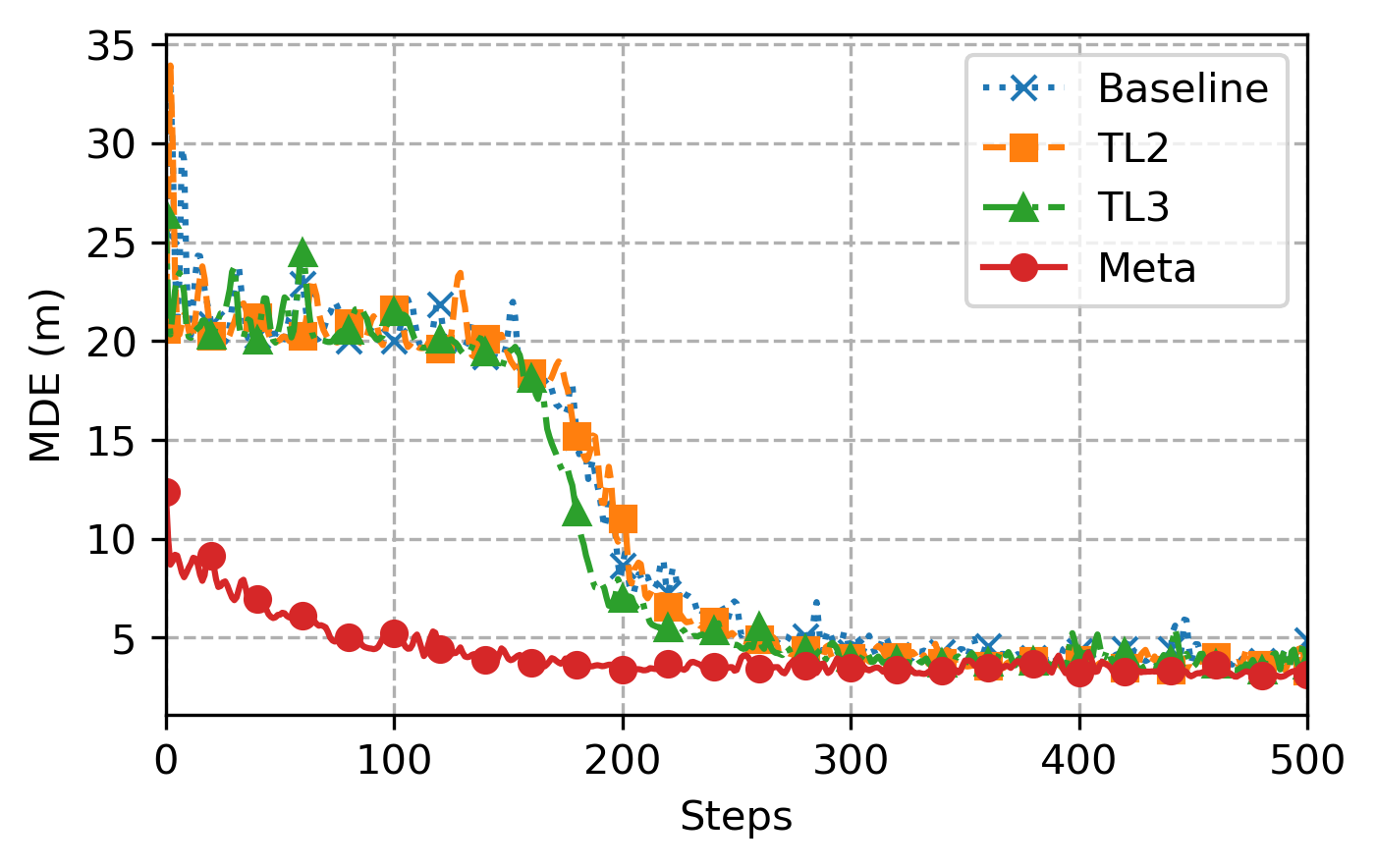}
        \caption{Meta-testing on floor 1}
        \label{fig:meta1}
    \end{subfigure}
    \caption{MetaGraphLoc performance evaluation in new environments.}
    \label{fig:meta-tesing}
\end{figure*}

In this section, we evaluate MetaGraphLoc's meta-learning approach by treating floors in the IMUWiFine dataset as independent environments. A subset of two floors ($K=2$ tasks) is used for meta-training (details in Algorithm \ref{alg:meta-training}). The remaining floor ($L=1$ task) is the meta-testing set to assess generalization to unseen environments. To account for potential bias in floor selection, we experiment with three configurations: (i) floors 1 \& 2 for meta-training, floor 3 for meta-testing; (ii) floors 1 \& 3 for meta-training, floor 2 for meta-testing; and (iii) floors 2 \& 3 for meta-training, floor 1 for meta-testing.

Within each configuration, the MetaGraphLoc meta-model is compared to three baselines: a GNN built from scratch (Baseline GNN), and two transfer learning models trained on single tasks ($TL_k$) (i.e, TL1 and TL2) from the meta-training set. Here, $k$ takes two values corresponding to each configuration's possible tasks.  Performance comparisons are presented in Fig. \ref{fig:meta-tesing}. Fig. \ref{fig:meta3} shows the MDE over 500 SGD steps for various models when tested on floor 3. The MetaGraphLoc model consistently outperforms the other models, achieving lower MDE throughout the steps. This indicates its superior ability to adapt and generalize to new environments compared to the baseline and transfer learning models. The baseline and TL models show higher and more varying MDE, especially in the initial steps, indicating less efficient learning and adaptation.

Fig. \ref{fig:meta2}  illustrates the MDE for the same set of models when tested on floor 2. Similar to the previous result, MetaGraphLoc demonstrates significantly lower MDE compared to the baseline GNN and TL models. Finally, Fig. \ref{fig:meta1} depicts the MDE for the models when tested on floor 1. The trend observed in the previous plots continues here. MetaGraphLoc achieves the lowest MDE across all steps, outperforming the baseline GNN and TL models. The other models show higher initial MDE and greater variability, further emphasizing the advantage of MetaGraphLoc's approach in adapting to unforeseen environments.

\subsection{Energy consumption and $CO_2$ footprint analysis}
TABLE~\ref{complexity_table} presents a detailed comparison of different GNN architectures for indoor localization, evaluated on training time, power consumption, CO2 footprint, inference time, and MDE. We used ECO2AI~\cite{budennyy2022eco2ai}, an open-source Python library that provides information on CPU and GPU consumption and estimates for corresponding CO2 emissions. The key findings are summarized as follows:

\subsubsection{Training time}
The MPNN exhibits the shortest training time at 210.76 seconds, while GraphSAGE (Sage) demonstrates the fastest inference time at 609.90 microseconds, highlighting their efficiency in training and real-time applications, respectively. Conversely, the GAT and DEC have the longest training times (369.32 and 342.33 seconds) and the highest power consumption (17.81 Wh for GAT and 22.07 Wh for DEC).
\subsubsection{Localization accuracy}
DEC achieves the lowest MDE at 2.69 meters, indicating superior localization accuracy. However, this comes at the expense of higher power consumption and inference time (5394.66 microseconds). On the other hand, GCN shows the highest MDE at 4.16 meters, suggesting it is the least accurate among the evaluated models.
\subsubsection{Environmental impact}
MPNN and Sage are the most environmental friendly approaches, both with a low CO2 footprint of 1.82 grams, whereas DEC incurs the highest CO2 emissions at 3.13 grams, reflecting its higher computational demands.
\subsubsection{Trade-offs and optimal choices}
MPNN and Sage offer a balanced trade-off, combining efficiency (short training time, low power consumption, low CO2 footprint, and fast inference) with reasonable accuracy (MDE of 3.43 m and 2.74 m, respectively). While models like MPNN and Sage provide lower computational overhead, DEC offers the highest accuracy. While DEC provides the highest accuracy, it requires significantly higher resources, making it suitable for scenarios where accuracy is paramount, and resource constraints are secondary. Overall, the choice of architecture involves a trade-off between accuracy, efficiency, power consumption, and environmental impact. 

\subsection{Future Work}
Based on the above findings, while the DEC model demonstrates superior localization accuracy, its computational overhead is notably higher compared to other architectures. As a consequence, deploying the DEC model in resource-constrained TinyML environments presents practical difficulties. The need for high computational power, extensive memory, and prolonged processing times can limit its application in edge devices or environments with limited hardware capabilities \cite{eldeeb2022traffic}. This increased complexity stems from the dynamic nature of the adjacency matrix, which requires additional computations during each training iteration.
To address these bottlenecks, several strategies can be explored:
\subsubsection*{Model compression and optimization}
 Techniques like model pruning, quantization, and knowledge distillation can be employed to reduce the size and computational complexity of the DEC model without significantly affecting its accuracy.
 
\subsubsection*{Hardware acceleration}
Utilizing specialized hardware, such as GPUs, TPUs, or dedicated edge AI chips, can help expedite both training and inference processes, thereby reducing power consumption and CO2 footprint \cite{8916327}.

\subsubsection*{Efficient algorithm design}
Developing more efficient algorithms for dynamic edge convolution that maintain accuracy while reducing computational overhead can lead to substantial improvements in performance metrics.

\subsubsection*{Hybrid approaches}
Combining DEC with other less resource-intensive models in a hierarchical or ensemble framework could balance the trade-offs between accuracy and efficiency.

In summary, while DEC offers impressive accuracy for indoor localization, addressing its bottlenecks in training time, power consumption, environmental impact, and inference time is crucial for broader and more sustainable deployment. Through targeted optimizations and innovative approaches, the performance of DEC can be enhanced to meet the demands of real-world applications.

\begin{table}[t!]
\centering
 \caption{The complexity evaluation of different deep learning architectures in terms of training time, power consumption, CO2 footprint, inference time and MDE.}
\label{complexity_table}
\begin{tabular}{@{}cccccc@{}}
\toprule
\textbf{Model} & \textbf{Train (s) }& \textbf{Power (Wh)} & \textbf{CO2 (g)}  & \textbf{Inference ($\boldsymbol\mu$s)} & \textbf{MDE}\\ \midrule
\midrule
GCN & $231.29$ & $14.16$ & $2.01$ & $1452.18$ & 4.16\\
\hline 
GAT & $369.32$ & $17.81$ & $2.53$ & $1483.94$ & 3.83\\
\hline 
MPNN & $210.76$ & $12.82$ & $1.82$ & $775.13$ & 3.43\\
\hline 
Cheb & $246.53$ & $14.09$ & $2.00$ & $1918.30$ & 2.85\\
\hline 
Sage & $220.78$ & $12.83$ & $1.82$ & $609.90$ & 2.74\\
\hline 
DEC & $342.33$ & $22.07$ & $3.13$ & $5394.66$ & $2.69$\\
\bottomrule
\end{tabular}
\end{table}

%% file: sections/conclusion.tex
\section{Conclusion}
\label{sec:conclusion}
This paper addressed the challenges of accurate indoor localization due to environmental variations and limited data availability. We proposed MetaGraphLoc, a novel system that leverages data fusion, GNNs with DEC, and meta-learning.
MetaGraphLoc demonstrated its effectiveness through several key findings. Data fusion significantly improves localization accuracy by combining WiFi RSSI measurements and IMU data. The GNN architecture with DEC effectively models indoor environments' spatial relationships and dynamics, outperforming static methods. Finally, the meta-learning framework facilitates fast adaptation to new environments, reducing data collection requirements.
These findings position MetaGraphLoc as a promising solution for indoor localization. This work contributes to the field of indoor localization by achieving improved accuracy, efficiency, and adaptability compared to existing approaches. This paves the way for more reliable and practical indoor location-based services in the ever-evolving IoT networks. Future research can further enhance our findings by exploring various directions, including integrating additional sensor modalities and addressing practical challenges such as computational complexity, hardware adaptability and real-time implementation.